\documentclass[jcp,aip,floatfix,preprintnumbers,superscriptaddress,preprint]{revtex4-1}
\usepackage{graphicx}
\usepackage{dcolumn}
\usepackage{bm}
\usepackage{pifont}
\usepackage{amsmath}
\usepackage{times}
\usepackage{xcolor}
\usepackage{epstopdf}

\begin{document}

\title{The effects of intrinsic noise on the behaviour of bistable cell regulatory systems
under quasi-steady state conditions}

\author{Roberto de la Cruz}

 \affiliation{Centre de Recerca Matem\`atica. Edifici C, Campus de Bellaterra, 08193 Bellaterra (Barcelona), Spain.}

 \affiliation{Departament de Matem\`atiques, Universitat Aton\`oma de Barcelona, 08193 Bellaterra (Barcelona), Spain.}

\author{Pilar Guerrero}

 \affiliation{Department of Mathematics, University College London, Gower Street, London WC1E 6BT, UK.}

 \author{Fabian Spill} 

 \affiliation{Department of Biomedical Engineering, Boston University, 44 Cummington Street, Boston MA 02215, USA.}

 \affiliation{Department of Mechanical Engineering, Massachusetts Institute of Technology, 77 Massachusetts Avenue, Cambridge, MA 02139, USA.}

 \author{Tom\'as Alarc\'on}

 \affiliation{Centre de Recerca Matem\`atica. Edifici C, Campus de Bellaterra, 08193 Bellaterra (Barcelona), Spain.}

 \affiliation{Departament de Matem\`atiques, Universitat Aton\`oma de Barcelona, 08193 Bellaterra (Barcelona), Spain.}

\date{\today}

\begin{abstract}
We analyse the effect of intrinsic fluctuations on the properties of bistable stochastic systems with time scale separation operating under1
quasi-steady state conditions. We first formulate a stochastic generalisation of the quasi-steady state approximation based on the
semi-classical approximation of the partial differential equation for the generating function associated with the Chemical Master Equation. Such approximation
proceeds by optimising an action functional whose associated set of Euler-Lagrange (Hamilton) equations provide the most likely fluctuation path. We show that, under
appropriate conditions granting time scale separation, the Hamiltonian can be re-scaled so that the set of Hamilton equations splits up into slow and
fast variables, whereby the quasi-steady state approximation can be applied. We analyse two particular examples of systems whose mean-field limit has
been shown to exhibit bi-stability: an enzyme-catalysed system of two mutually-inhibitory proteins and a gene regulatory circuit with self-activation.
Our theory establishes that the number of molecules of the conserved species are order parameters whose variation regulates bistable
behaviour in the associated systems beyond the predictions of the mean-field theory. This prediction is fully confirmed by direct numerical
simulations using the stochastic simulation algorithm. This result allows us to propose strategies whereby, by varying the number
of molecules of the three conserved chemical species, cell properties associated to bistable behaviour (phenotype, cell-cycle status, etc.) can be controlled.
\end{abstract}

\maketitle

\section{Introduction}

The networks of interacting genes and proteins that are responsible for regulation, signalling and response, and which, ultimately, orchestrate cell
function, are under the effect of noise \cite{kepler2001,kaern2005,maheshri2007,losick2008,raj2008}. This randomness materialises in the form of fluctuations of the number of molecules of the species involved, subsequently leading to fluctuations in their activity.  Besides external perturbations, biochemical reactions can be intrinsically noisy, especially when the number of molecules is very low. 

Far from necessarily being a mere disturbance, fluctuations are an essential component of the dynamics of cellular regulatory systems which, in many
instances, are exploited to improve cell function \cite{cai2008,eldar2010}. For example, randomness has been shown to enhance the ability of cells to
adapt and increase their fitness in random or variable environments \cite{kussell2005,acar2008,guerrero2015}. Random noise also serves the purpose of
assisting cell populations to sustain phenotypic variation by enabling cells to explore the phase space
\cite{maheshri2007,raj2008,losick2008,macarthur2008,eldar2010,balazsi2011}.  

One of the mechanisms that allows noise-induced phenotypic variability relays on multi-stability \cite{cinquin2005,jaeger2014}. The basis of this
mechanism was first proposed by Kauffman \cite{kauffman1993}, who associated phenotypes or differentiated states to the stable attractors of the
dynamical systems associated to gene and protein interaction networks. In the presence of noise, the corresponding phase space generates an epigenetic
landscape, where cells exposed to the same environment and signalling cues coexist
in different cellular phenotypes \cite{huang2012}.   

Multi-stability is also an essential element in the control of cell response and function via signalling pathways \cite{tyson2003}. In particular,
bi-stability as a means to generate reliable switching behaviour is widely utilised in numerous pathways such as the
apoptosis \cite{legewie2006}, cell survival \cite{legewie2007}, differentiation \cite{kalmar2009}, and cell-cycle progression
\cite{ferrell2001,tyson2001} pathways. For example, bi-stability is used to regulate such critical cell functions such as the transition from
quiescence to proliferation through
bistable behaviour associated with the Rb-E2F switch within the regulatory machinery of the mammalian cell-cycle
\cite{gerard2009,gerard2012,yao2008,yao2014,gerard2014,bedessem2014}. 

A common theme which appears when trying to model cell regulatory systems is separation of time scales, i.e. the presence of multiple processes
evolving on widely diverse time scales. When noise is ignored and systems are treated in terms of deterministic mean-field descriptions, such
separation of time scales and the associated slow-fast dynamics are often exploited for several forms of model reduction, of which one of the most
common is the so-called quasi-steady state approximation (QSSA) \cite{keener1998}. This approximation is ubiquitously used whenever regulatory
processes involve enzyme catalysis, which is a central regulation mechanism in cell function \cite{tyson2003}. In this paper, we investigate the
effects of intrinsic noise on the bi-stability of two particular systems, namely, an enzyme-catalysed system of mutual inhibition and a gene
regulatory
circuit with self-activation. The mean-field limit of both these systems has been shown to exhibit bi-stability \cite{tyson2001,frigola2012}. The aim of this paper is to analyse how noise alters the mean-field behaviour associated to these systems when they operate under quasi-steady state conditions. 

 We note that this work does not concern the subject of noise-induced bifurcations \cite{garciaojalvo1999}. Such phenomenon has been studied in many situations, including biological systems. An example which is closely related to the systems we analyse here is the so-called enzymatic futile cycles. Samoilov et al. \cite{samoilov2005} have shown that noise associated to the number of enzymes induce bistability. In the absence of this source of noise, i.e. in the mean-field limit, the system does not exhibit bistable behaviour. The treatment of this phenomena would require to go to higher orders in the WKB expansion, which we do not explore here.

The issue of separation of time scales in stochastic models of enzyme catalysis has been addressed using a number of different approaches. Several
such analysis have been carried out in which the QSSA is directly applied to the master equation by setting the fast
reactions in partial equilibrium (i.e. the probability distribution corresponding to the fast variables remains unchanged), and letting the
rest of the system to evolve according to a reduced stochastic dynamic \cite{rao2003,turner2004}. Other approaches have been proposed such as the QSSA
to the exact Fokker-Planck equation that can be derived from the Poisson representation of the chemical master equation \cite{thomas2010}. Approaches
based on enumeration techniques have also been formulated \cite{doka2012}. Furthermore, Thomas et al.\cite{thomas2012} have recently formulated a
rigorous method to eliminate fast stochastic variables in monostable systems using projector operators within the linear noise approximation
\cite{thomas2012}. Methods for model reduction based on perturbation analysis have been developed in \cite{alarcon2014,bruna2014}. Additionally,
driven by the need of more efficient numerical methods, there has been much activity regarding the development of numerical methods for stochastic
systems with multiple time-scales \cite{burrage2004,vlachos2005,macnamara2008}. Several of these methods are variations of the stochastic 
simulation algorithm \cite{rao2003,cao2005,cao2005b,samant2005,e2007b,sanft2011} or the $\tau$-leap method \cite{rathinam2003} where the existence of
fast and
slow variables is exploited to enhance their performance with respect to the standard algorithms. Another family of such numerical methods is that of
the so-called hybrid methods, where classical deterministic rate equations or stochastic Langevin equations for the fast variables are combined with
the classical stochastic simulation algorithm for the slow variables \cite{haseltine2002,salis2005}.  Other related methods were studied in \cite{assaf2006,newby13,kang2013}.

Here, we advance the formalism developed in \cite{alarcon2014}, in which a method based on the semi-classical approximation of the Chemical Master
Equation allows to evaluate the effects of intrinsic random noise under quasi-steady conditions. In our analysis of the Michaelis-Menten model of
enzyme catalysis in \cite{alarcon2014}, we showed that the semi-classical quasi-steady state approximation reveals that the velocity of the enzymatic
reaction is modified with respect to the mean-field estimate by a quantity which is proportional to the total number of molecules of the (conserved)
enzyme. In this paper, we extend this formalism to show that, associated to each conserved molecular species, the associated (constant) number of
molecules is a bifurcation parameter which can drive the system into bi-stability beyond the predictions of the mean-field theory. We then proceed to
test our theoretical results by means of direct numerical simulation of the Chemical Master Equation using the stochastic simulation algorithm
\cite{gillespie1976}. We should note the Hamiltonian formalism derived from the semi-classical approximation is formulated on a 
continuum of particles, which requires the number of particles to be large enough. This must hold true for all the species in our model, 
both fast and slow. Since this separation between fast and slow species is based on their relative abundance, one must be careful that the 
scaling assumptions are consistent, particularly in the case of the model of self-activating gene regulatory circuit where the 
number of binding sites is typically small. This assumption, however, has been used in previous studies \cite{assaf2011}. Also we show that our 
simulation results of the full stochastic processes agree with our analysis and, therefore, our re-scaled equations are able to predict 
the behaviour of the system. We note that the mean-field limit, which is conventionally obtained by 
ignoring noise in the limit of large particle numbers, is obtained by setting the momenta in our phase-space formalism to $1$.

The approximation we develop in this paper falls within the general framework of the optimal fluctuation path theory \cite{bressloff2014a}. This framework is a particular case of the large deviation theory which allows us to study rare events (i.e. events whose frequency is exponentially small with system size). Within these framework we will show that, upon carrying out the QSSA, the only source of noise in the system is associated to the random initial conditions of the species whose numbers are conserved. We therefore predict that a population of cells, each having a random number of  conserved molecules, will have a bimodal distribution.

This paper is organised as follows. Section 2 is devoted to a detailed exposition of the semi-classical quasi-steady state approximation for
stochastic systems. In Sections 3 and 4, we apply this formalism to analyse the behaviour of a bistable enzyme-catalysed system and a gene regulatory
circuit of auto-activation, respectively. We will show that our semi-classical quasi-steady state theory allows us to study the effect of intrinsic
noise on the behaviour of these systems beyond the predictions of their mean-field descriptions. We also verify our theoretical predictions by means
of direct stochastic simulations. Finally in Section 5, we summarise our results and discuss their relevance.  

\section{Semi-classical quasi-steady state approximation}\label{sec:theory}

Our aim in this paper is to formulate a stochastic generalisation of the quasi-steady state approximation for enzyme-catalysed reactions and
simple circuits of gene regulation and use such approximation to determine if the presence of noise has effects on the behaviour of the system beyond
the predictions of the corresponding mean-field models. Specifically, we analyse stochastic systems for which the mean-field models predicts
bi-stability and investigate how such behaviour is affected by stochastic effects. Our analysis is carried out in the
context of Markovian models of the corresponding reaction mechanisms formulated in terms of the so-called chemical master
equation (CME) \cite{vankampen2007}. Two example of such stochastic systems, a bistable enzyme-catalysed system and a gene regulatory circuit of
auto-activation, are formulated and analysed in detail in Sections \ref{sec:enzyme} and \ref{sec:gene}, respectively. Following \cite{alarcon2014}, we
formulate the QSS approximation for the asymptotic solution of the CME obtained by means of large deviations/WKB approximations
\cite{kubo1973,alarcon2007,touchette2009}. The CME is given:

\begin{equation}\label{eq:cme}
\frac{\partial P(X,t)}{\partial t}=\sum_i(W_i(X-r_i)P(X-r_i,t)-W_i(X)P(X,t))
\end{equation}

\noindent where $W_i(X)$ is the transition rate corresponding to reaction channel $i$ and $r_i$ is a vector whose entries denote the change in the
number of molecules of each molecular species when reaction channel $i$ fires up, i.e. $P(X(t+\Delta t)=X(t)+r_i\vert x(t))=W_i(X)\Delta
t$. 

An alternative way to analyse the dynamics of continuous-time Markov processes on a discrete space of states is to derive an equation for the
generating function, $G(p_1,\dots,p_n,t)$ of the corresponding probabilistic density:

\begin{equation}
G(p_1,\dots,p_n,t)=\sum_x p_1^{X_1}p_2^{X_2}\cdots p_n^{X_n}P(X_1,\dots,X_n,t)
\end{equation}

\noindent where $P(X_1,\dots,X_n,t)$ is the solution of the Master Equation (\ref{eq:cme}). $G(p_1,\dots,p_n,t)$ satisfies a partial differential
equation (PDE) which can be derived from the Master Equation. This PDE is the basic element of the so-called momentum representation of the Master
Equation \cite{doi1976,peliti1985,assaf2006,assaf2010,kang2013}. 

Although closed, analytic
solutions are rarely available, the PDE for the generating function admits a perturbative solution, which is commonly obtained by means of the WKB
method \cite{assaf2010}. More specifically, the (linear) PDE that governs the evolution of the generating function can be written as:

\begin{equation}\label{eq:charfuncPDE}
\frac{\partial G}{\partial t}=H_k\left(p_1,\dots,p_n,\partial_{p_1},\dots,\partial_{p_n}\right)G(p_1,\dots,p_n,t) 
\end{equation}

\noindent where the operator $H_k$ is determined by the reaction rates of the Master Equation (\ref{eq:cme}). Furthermore, the solution to this
equation
must satisfy the normalisation condition $G(p_1=1,\dots,p_n=1,t)=1$ for all $t$. This PDE, or, equivalently, the operator $H$, are obtained by
multiplying both sides of the Master Equation (\ref{eq:cme}) by $\prod_{i=1}^{n}p_i^{X_i}$ and summing up over all the possible values of
$(X_1,\dots,X_n)$ 

From the mathematical point of view, Eq. (\ref{eq:charfuncPDE}) is a Schr\"odinger-like equation and, therefore, there is a plethora of methods at our
disposal in order to analyse it. In particular, when the fluctuations are (assumed to be) small, it is common to resort to WKB methods
\cite{kubo1973,alarcon2007,gonze2002}. This approach is based on the WKB-like Ansatz that $G(p_1,\dots,p_n,t)=e^{-S(p_1,\dots,p_n,t)}$. By substituting this Ansatz in Eq.
(\ref{eq:charfuncPDE}) we obtain the following Hamilton-Jacobi equation for the function $S(p_1,\dots,p_n,t)$:

\begin{equation}\label{eq:hamjac}
\frac{\partial S}{\partial t}=-H_k\left(p_1,\dots,p_n,\frac{\partial S}{\partial p_1},\dots,\frac{\partial S}{\partial p_n}\right)
\end{equation}

Instead of directly tackling the explicit solution of Eq. (\ref{eq:hamjac}), we will use the so-called semi-classical approximation. We use the Feynman
path-integral representation which yields a solution to Eq. (\ref{eq:charfuncPDE}) of the type
\cite{peliti1985,feynman2010,kubo1973,dickman2003,elgart2004,tauber2005}:

\begin{equation}\label{eq:pathintegral}
G(p_1,\dots,p_n,t)=\int_0^t e^{-S(p_1,\dots,p_n,Q_1,\dots,Q_n)}{\cal D}Q(s){\cal D}p(s),
\end{equation}

\noindent where ${\cal D}Q(s){\cal D}p(s)$ indicates integration over the space of all possible trajectories and
$S(p_1,\dots,p_n,Q_1,\dots,Q_n)$ is given by \cite{kubo1973}:

\begin{eqnarray}\label{eq:actionintegral}
\nonumber S(p_1,\dots,p_n,Q_1,\dots,Q_n)=&&-\int_0^t\left(H_k(p_1,\dots,p_n,Q_1,\dots,Q_n)+\sum_{i=1}^nQ_i(s)\dot{p}_i(s)\right)ds \\
&& + \sum_{i=1}^nS_{0,i}(p_i,Q_i),
\end{eqnarray}

\noindent where the position operators in the momentum representation have been defined as $Q_i\equiv \partial_{p_i}$ with the commutation relation
$[Q_i,p_j]=S_{0,i}\delta_{i,j}$. $S_{0,i}(p_i,Q_i)$ corresponds to the action associated with the generating function of the probability
distribution function of the initial value of each variable, $X_i(t=0)$, which are assumed to be independent random variables.

The so-called semi-classical approximation consists of approximating the path integral in Eq. (\ref{eq:pathintegral}) by

\begin{equation}\label{eq:geomopt}
G(p_1,\dots,p_n,t)=e^{-S(p_1,\dots,p_n,t)}
\end{equation}

\noindent where $p_1(t),\dots,p_n(t)$ are now the solutions of the Hamilton equations, i.e. the orbits which maximise the action $S$:

\begin{eqnarray}
\label{eq:hameqs-p} && \frac{dp_i}{dt}=-\frac{\partial H_k}{\partial Q_i}\\
\label{eq:hameqs-q} && \frac{dQ_i}{dt}=\frac{\partial H_k}{\partial p_i} 
\end{eqnarray}

\noindent where the pair ($Q_i$,$p_i$) are the generalised coordinates corresponding to chemical species $i=1,\dots,n$. These
equations are (formally) solved with boundary conditions\cite{elgart2004} $Q_i(0)=x_{i}(0)$, $p_i(t)=p_i$, where $x_i(0)$ is the initial number of
molecules of species $i$.

Eqs. (\ref{eq:hameqs-p})-(\ref{eq:hameqs-q}) are the starting point for the formulation of the semi-classical quasi-steady state approximation
(SCQSSA) \cite{alarcon2014}. In order to proceed further, we assume, as per the Briggs-Haldane treatment of the Michealis-Menten model for enzyme
kinetics \cite{briggs1925,keener1998}, that the species involved in the system under scrutiny are divided into two groups according to their
characteristic scales. More specifically, we have a subset of chemical species whose numbers, $X_i$, scale as:

\begin{equation}
X_i=Sx_i, 
\end{equation}

\noindent where $x_i=O(1)$, whilst the remaining species are such that their numbers, $X_j$, scale as:

\begin{equation}
X_j=Ex_j, 
\end{equation} 

\noindent where $x_j=O(1)$. Key to our approach is the fact that $S$ and $E$ must be such that:

\begin{equation}\label{eq:epsilon}
\epsilon=\frac{E}{S}\ll 1. 
\end{equation}

\noindent We further assume that the generalised coordinates, $Q_i$, scale in the same fashion as the corresponding variable $X_i$, i.e.

\begin{equation}
Q_i=Sq_i, 
\end{equation}

\noindent where $q_i=O(1)$. We refer to the variables belonging to this subset as \emph{slow variables}. Similarly,

\begin{equation}
Q_j=Eq_j, 
\end{equation}  
 
\noindent where $q_j=O(1)$, which are referred to as \emph{fast variables}. Moreover, we assume that the moment coordinates, $p_i$, are all
independent of $S$ and $E$, and therefore remain invariant under rescaling.

Under this scaling for the generalised coordinates, we define the following scale transformation for the Hamiltonian in Eq. (\ref{eq:actionintegral}):

\begin{equation}\label{eq:scaledham}
H_k(p_1,\dots,p_n,Q_1,\dots,Q_n)=k_JS^kE^lH_{\kappa}(p_1,\dots,p_n,q_1,\dots,q_n)
\end{equation}

\noindent where $J$ identifies the reaction with the largest order among all the reactions that compose the dynamics and $k_J$ is the
corresponding rate constant. For example, in the case of the bistable enzyme-catalysed system whose reactions or elementary events and the
corresponding transition rates are given in Table \ref{tab:smm-bistable}, $J=1$, as this reaction is order 3 whereas all the others are order 0, 1, or
2. In the case of the self-activating gene regulatory circuit, Table \ref{tab:smm-gene}, $J=3$, since this reaction is order 3 whereas the remaining
ones are order 1 at most. The exponents $k$ and $l$ correspond to the number of slow and fast variables involved in the transition rate $W_{J}$, respectively.

The last step is to rescale the time variable so that a dimensionless variable, $\tau$, is defined such that:

\begin{equation}\label{eq:scaledtime}
\tau=k_JS^{k-1}E^lt
\end{equation}

It is now a trivial exercise to check that, upon rescaling, Eqs. (\ref{eq:hameqs-p})-(\ref{eq:hameqs-q}) read

\begin{eqnarray}
\label{eq:hameqs-p-slow} && \frac{dp_i}{d\tau}=-\frac{\partial H_{\kappa}}{\partial q_i},\\
\label{eq:hameqs-q-slow} && \frac{dq_i}{d\tau}=\frac{\partial H_{\kappa}}{\partial p_i}, 
\end{eqnarray}

\noindent for the slow variables. By contrast, rescaling of the Hamilton equations corresponding to the subset of fast variables leads to:

\begin{eqnarray}
\label{eq:hameqs-p-fast} && \epsilon\frac{dp_j}{d\tau}=-\frac{\partial H_{\kappa}}{\partial q_j},\\
\label{eq:hameqs-q-fast} && \epsilon\frac{dq_j}{d\tau}=\frac{\partial H_{\kappa}}{\partial p_j}, 
\end{eqnarray} 

\noindent where $\epsilon$ is defined in Eq. (\ref{eq:epsilon}). The QSS approximation consists on assuming that $\epsilon\frac{dp_j}{d\tau}\simeq 0$
and $\epsilon\frac{dq_j}{d\tau}\simeq 0$ in Eqs. (\ref{eq:hameqs-p-fast})-(\ref{eq:hameqs-q-fast}), 

\begin{eqnarray}
\label{eq:hameqs-p-QSSA} && -\frac{\partial H_{\kappa}}{\partial q_j}=0,\\
\label{eq:hameqs-q-QSSA} && \frac{\partial H_{\kappa}}{\partial p_j}=0, 
\end{eqnarray} 

\noindent resulting in a differential-algebraic system of
equations which provides us with the semi-classical quasi-steady state approximation (SCQSSA).

\section{Bistable enzyme-catalysed systems}\label{sec:enzyme} 

\begin{figure}
\begin{center}
\includegraphics[scale=0.5]{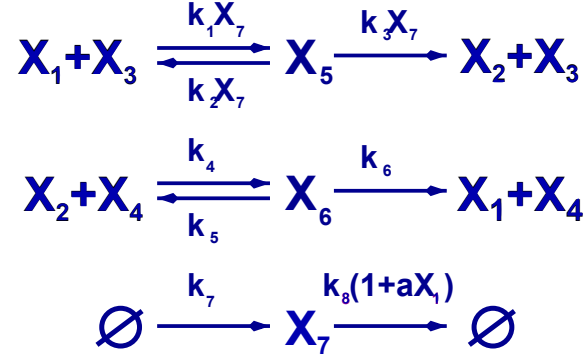} 
\caption{Reactions for the bistable enzyme-catalysed system proposed by Tyson \& Novak \cite{tyson2001}. $X_1$ represents active Cdh/Apc, $X_2$
inactive Cdh/Apc, $X_3$ inactivating enzymes, $X_4$ activating enzymes, $X_5$ active Cdh/Apc-inactivating-enzyme complexes,
$X_6$ inactive Cdh/Apc-activating-enzyme complexes, and $X_7$ the number of CycB-CDK complexes. The first two reactions correspond to
enzyme-catalysed inactivation and activation of Cdh/APC. The third reaction corresponds to the dynamics of CycB activity: synthesis at a constant
rate, $k_7$, and degradation by natural decay and active Cdh/Apc-induced inactivation.}\label{fig:bistable-mm}
\end{center}
\end{figure}

\begin{table}[htb]
\begin{center}
\begin{tabular}{ll}
Variable & Description \\\hline
$X_1$, $X_2$ & Number of active and inactive (respectively) Cdh1 molecules\\
$X_3$, $X_4$ & Number of Cdh1-inactivating and Cdh1-activating (respectively) enzyme molecules\\
$X_5$, $X_6$ & Number of enzyme-active Cdh1 and enzyme-inactive Cdh1 (respectively) complexes\\
$X_7$, & Number of active cyclin molecules \\\hline
\end{tabular} 
\begin{tabular}{lll}
Transition rate & r & Event \\\hline
$W_1(x)=k_{1}X_7X_1X_3$ & $r_1=(-1,0,-1,0,+1,0,0)$ & Enzyme and active Cdh1 form complex\\
$W_2(x)=k_{2}X_7X_5$ & $r_2=(+1,0,+1,0,-1,0,0)$ & Enzyme-active Cdh1 complex splits\\
$W_3(x)=k_{3}X_7X_5$ & $r_3=(0,+1,+1,0,-1,0,0)$ & Inactivation of Cdh1 and enzyme release\\
$W_4(x)=k_{4}X_2X_4$ & $r_4=(0,-1,0,-1,0,+1,0)$ & Enzyme and inactive Cdh1 form complex\\
$W_5(x)=k_{5}X_6$ & $r_5=(0,+1,0,+1,0,-1,0)$ & Enzyme-inactive Cdh1 complex splits\\
$W_6(x)=k_{6}X_6$ & $r_6=(+1,0,0,+1,0,-1,0)$ & Activation of Cdh1 and enzyme release \\
$W_7(x)=k_{7}$ & $r_7=(0,0,0,0,0,0,+1)$ & CycB synthesis\\
$W_8(x)=k_{8}(1+aX_1)X_7$ & $r_8=(0,0,0,0,0,0,-1)$ & CycB degradation
\end{tabular} 
\end{center}
\caption{Random variables and transition rates of the stochastic model associated to the enzymatic reaction shown in Fig.
\ref{fig:bistable-mm}.}\label{tab:smm-bistable}
\end{table}

As a prototype of a bistable enzyme-catalysed system, we analyse a stochastic system proposed in \cite{alarcon2014,guerrero2014a}, whose mean-field
limit has been
shown to correspond to a bistable system which is a part of a model for the G$_1$/S transition of the eukaryote cell cycle proposed in
\cite{tyson2001}. Tyson \& Novak \cite{tyson2001} have formulated a (deterministic) model of the cell cycle such that the core of the system
regulating the G$_1$/S transition is a system of two mutually-repressing proteins (Cdh1 and CycB). This system of mutual repression
gives rise to a bistable system where one of the stable steady states is identified with the G$_1$ phase whereas the other corresponds to a state
where
the cell is ready to go through the other three phases of the cell-cycle, known as S, G$_2$, and M. This central module, which is the one we focus
on, is acted upon by a complex regulatory network which monitors if conditions are met for the cell to undergo this transition and accounts for its
accurate timing. Presently, we ignore this network and focus on the central bistable system. It is shown in \cite{tyson2001} that the mean field
version of the model exhibits bistable behaviour as a function of a bifurcation parameter $m$, i.e. the mass of the cell. For very small values of
$m$, the system is locked into a high (low) Cdh1(CycB)-level stable fixed point (i.e. into the G$_1$ phase). For very large values $m$, the system has
only one stable steady state corresponding to a low (high) Cdh1(CycB)-level fixed point. For intermediate values of $m$ the system exhibits bistability, i.e. both of these stable fixed
points coexist with an unstable saddle point. In this section, we focus on how noise alters the behaviour of the mean-field dynamics. 

The transition rates corresponding to the different reactions
involved in the stochastic model associated to the enzyme-regulated kinetics shown in Fig. \ref{fig:bistable-mm} are given in Table
\ref{tab:smm-bistable}. This kinetics corresponds to the enzyme regulated activation and inhibition of Cdh1 (an inhibitor of cell-cycle progression).
Cdh1 inactivation is further (up)regulated by the presence of CycB, an activator of cell-cycle progression. CycB is synthesised and degraded at basal
rates and is further degraded in the presence of active Cdh1 (see Fig. \ref{fig:bistable-mm}). Therefore, the resulting dynamics leads to
a system with mutual inhibition which produces bistable behaviour. It is important to note that the associated reaction kinetics exhibits three
conservation laws (see Table \ref{tab:smm-bistable}): $X_3+X_5=e_0$, $X_4+X_6=e_0$, and $X_1+X_2+X_5+X_6=s_0$. The first two of these conservation
laws are associated to the conservation of the number of Cdh1-inhibiting and Cdh1-activating enzymes, respectively, whilst the latter expresses the
conservation of the total number of Cdh1 molecules. The quantities $e_0$ and $s_0$ are the (conserved) number of enzymes and Cdh1, respectively. Note
that, as per the methodology developed in Section \ref{sec:theory}, we assume that $s_0=O(S)$ and $e_0=O(E)$.

\begin{table}[htb]
\begin{center}
\begin{tabular}{lll}
Rescaled variables & \vline \mbox{ } & Dimensionless parameters\\\hline 
$\tau=k_1ESt$ & \vline \mbox{ }  & $\epsilon=E/S$, $\alpha=aS$\\
$q_1=Q_1/S$ & \vline \mbox{ }  & $\kappa_2=k_2/(k_1S)$ \\
$q_2=Q_2/S$ & \vline \mbox{ }  & $\kappa_3=k_3/(k_1S)$\\
$q_3=Q_3/E$ & \vline \mbox{ }  & $\kappa_4=k_4/(k_1S)$\\
$q_4=Q_4/E$ & \vline \mbox{ }  & $\kappa_5=k_5/(k_1S^2)$  \\
$q_5=Q_5/E$ & \vline \mbox{ }  & $\kappa_6=k_6/(k_1S^2)$  \\
$q_6=Q_6/E$ & \vline \mbox{ }  & $\kappa_7=k_7/(k_1ES^2)$ \\
$q_7=Q_7/S$ & \vline \mbox{ }  & $\kappa_8=k_8/(k_1ES)$
\end{tabular} 
\end{center}
\caption{Dimensionless variables used in Eqs. (\ref{eq:kappaham}). $S$ and
$E$ are the
average concentration of Cdh1 (active plus inactive) and the average concentration of both Cdh1-activating and Cdh1-inactivating enzymes, respectively. We
further assume that the stationary concentration of
active CycB also scales with $S$.}\label{tab:dless-tn}
\end{table}

The corresponding stochastic Hamiltonian,  $H_k$, which is derived by applying the methodology of Section \ref{sec:theory} to the Master Equation
associated to the chemical kinetics described in Table \ref{tab:smm-bistable}, can be split into three parts,

\begin{equation}\label{eq:mmb2}
H_k(p_1,\dots,p_7,Q_1,\dots,Q_7)=H_A+H_I+H_B, 
\end{equation}

\noindent where $H_I$ is the Hamiltonian corresponding to the  CycB-regulated enzymatic inactivation of Cdh1 (reactions 1 to 3 in Table \ref{tab:smm-bistable}):

\begin{equation}\label{eq:mmb5}
H_I(p,Q)=k_4(p_6-p_2p_4)Q_2Q_4+k_5(p_2p_4-p_6)Q_6+k_6(p_1p_4-p_6)Q_6,
\end{equation}

\noindent $H_A$ corresponds to enzymatic activation of Cdh1 (reactions 4 to 6 in Table \ref{tab:smm-bistable}):

\begin{equation}\label{eq:mmb3}
H_A(p,Q)=k_1p_7(p_5-p_1p_3)Q_1Q_3Q_7+k_2p_7(p_1p_3-p_5)Q_5Q_7+k_3p_7(p_2p_3-p_5)Q_5Q_7, 
\end{equation}

\noindent and, finally, $H_B$, which corresponds to synthesis and degradation of CycB, is given by (reactions 7 and 8 in Table \ref{tab:smm-bistable}):

\begin{equation}\label{eq:mmb4}
H_B(p,Q)=k_7(p_7-1)+k_8(1-p_7)Q_7+k_8ap_1(1-p_7)Q_1Q_7.
\end{equation}

We now proceed to apply the procedure explained in Section 2 in order to obtain the SCQSSA for the system determined by the transition rates given in
Table \ref{tab:smm-bistable}. We first need to determine which of the variables are slow variables and which ones are fast variables. As shown in
Table \ref{tab:dless-tn}, the pairs $(p_1,Q_1)$, $(p_2,Q_2)$, and $(p_7,Q_7)$, corresponding to the active and inactive forms of Cdh1 and to CycB,
respectively, are the slow generalised coordinates, as the generalised positions scale with $s_0$. The remaining generalised coordinates scale as
$e_0$ and, therefore, are fast variables. Furthermore, the rescaled Hamiltonian is given by:

\begin{equation}\label{eq:resham}
H_k(p,Q)=k_1ES^2H_{\kappa}(p,q)
\end{equation}

\noindent where

\begin{equation}
H_{\kappa}(p,q)=H_{\kappa,A}+H_{\kappa,I}+H_{\kappa,B}, 
\end{equation}

\noindent with

\begin{eqnarray}\label{eq:kappaham}
\nonumber && H_{\kappa,I}=\kappa_4(p_6-p_2p_4)q_2q_4+\kappa_5(p_2p_4-p_6)q_6+\kappa_6(p_1p_4-p_6)q_6\\
\nonumber && H_{\kappa,A}=p_7(p_5-p_1p_3)q_1q_3q_7+\kappa_2p_7(p_1p_3-p_5)q_5q_7+\kappa_3p_7(p_2p_3-p_5)q_5q_7\\
&& H_{\kappa,B}=\kappa_7(p_7-1)+{\kappa}_8(1-p_7)q_7+\kappa_8\alpha p_1(1-p_7)q_1q_7
\end{eqnarray}

\noindent The rescaled parameters $\kappa_i$ are given in Table \ref{tab:dless-tn}. Last, by rescaling time and defining the dimensionless time
variable as $\tau=k_1ESt$ (Table \ref{tab:dless-tn}), the SCQSSA equations (\ref{eq:hameqs-p-slow})-(\ref{eq:hameqs-q-slow}) and
(\ref{eq:hameqs-p-QSSA})-(\ref{eq:hameqs-q-QSSA}) lead to (see \cite{alarcon2014} for a detailed derivation):

\begin{eqnarray}
&& \frac{dq_1}{d\tau}=p_4p_{e_4}\frac{\kappa_6q_2}{q_2+J_2}-p_7p_3p_{e_3}\frac{\kappa_3q_7q_1}{q_1+J_1}+\kappa_8\alpha(1-p_7)q_7q_1\\
&& \frac{dq_2}{d\tau}=-p_4p_{e_4}\frac{\kappa_6q_2}{q_2+J_2}+p_7p_3p_{e_3}\frac{\kappa_3q_7q_1}{q_1+J_1}\\
&& \frac{dq_7}{d\tau}=\kappa_7-\kappa_8(1+\alpha p_1q_1)q_7\\
&& p_5=p_3p_1\\ 
&& p_6=p_4p_2\\
&& \frac{dp_7}{d\tau}=-(1-p_7)\kappa_8(1+\alpha p_1q_1)
\end{eqnarray}

\noindent where $p_1=p_2$, $p_3$, and $p_4$ are constants to be determined and $J_1=\kappa_2+\kappa_3$ and $J_2=\kappa_4^{-1}(\kappa_5+\kappa_6)$, and
$p_{e_3}=e_3/E$ and $p_{e_4}=e_4/E$. Note that for $q_1(\tau)+q_2(\tau)=p_c$, with $p_c=s_0/S$, to hold $p_7=1$ must be satisfied. In this case, we have

\begin{eqnarray}
\label{eq:hjqss-x1} && \frac{dq_1}{d\tau}=p_4p_{e_4}\frac{\kappa_6(p_c-q_1)}{(p_c-q_1)+J_2}-p_3p_{e_3}\frac{\kappa_3mq_7q_1}{q_1+J_1}\\
\label{eq:hjqss-x7} && \frac{dq_7}{d\tau}=\kappa_7-\kappa_8(1+\alpha p_1q_1)q_7\\
\label{eq:hjqss-p3} && p_5=p_3p_1\\ 
\label{eq:hjqss-p4} && p_6=p_4p_1
\end{eqnarray}

As shown in \cite{alarcon2014}, the parameter values are determined by comparing the corresponding mean-field approximation, which is obtained by
taking $p_i=1$ \cite{elgart2004}, and $p_c=p_{e_3}=p_{e_4}=1$, i.e. the total number of molecules of Cdh1 and its activating and inhibiting enzymes be
exactly equal to its average, i.e. $s_0=S$ and $e_3=e_4=E$, to the system originally proposed by Tyson \& Novak \cite{tyson2001}. In Eq.
(\ref{eq:hjqss-x1}) we have redefined
$\kappa_3\to \kappa_3m$ in order to make explicit the dependence on the bifurcation parameter, $m$, as used by Tyson \& Novak \cite{tyson2001}. The
parameter values are shown in Table \ref{tab:par-val}.

\begin{table}[htb]
\begin{center}
\begin{tabular}{llll}
Rescaled parameter & \vline Parameter & \vline \mbox{Units} & \vline Reference\\\hline 
$\kappa_2=J_4-\kappa_3$ & \vline $a_1'=0.04$ & \vline \mbox{min}$^{-1}$ & \vline   \cite{tyson2001} \\
$\kappa_3m=\frac{a_4m}{k_1ES}$ & \vline $a_2'=0.04$ & \vline \mbox{min}$^{-1}$ & \vline   \cite{tyson2001} \\
$\kappa_6=\frac{a_3'}{k_1ES}$ & \vline $a_2''=1$ & \vline \mbox{min}$^{-1}$ & \vline   \cite{tyson2001} \\
$\kappa_5=\kappa_4J_3-\kappa_6$ & \vline $a_3=1$ & \vline \mbox{min}$^{-1}$ & \vline   \cite{tyson2001} \\
$\kappa_7=\frac{a_1'}{k_1ES}$ & \vline $a_4=35$ & \vline \mbox{min}$^{-1}$ & \vline   \cite{tyson2001}\\
$\kappa_8=\frac{a_2'}{k_1ES}$ & \vline $m=0.3$ & \vline Dimensionless & \vline   -- \\
$a=\frac{a_2''}{k_1ES\kappa_8}$ & \vline $E=0.01$ & \vline Dimensionless & \vline   \cite{alarcon2014} \\
\mbox{} & \vline $S=1.0$ & \vline Dimensionless & \vline   \cite{alarcon2014} \\
\mbox{}  & \vline $k_1=1$ & \vline \mbox{min}$^{-1}$ & \vline   \cite{rao2003}  \\
\mbox{}  & \vline $\kappa_4=\kappa_3$ & \vline Dimensionless & \vline \cite{alarcon2014} \\
\mbox{}  & \vline $J_3=J_4=0.04$ & \vline Dimensionless & \vline \cite{tyson2001} \\
\end{tabular} 
\end{center}
\caption{Parameter values used in simulations of the stochastic bistable enzyme-catalysed system}\label{tab:par-val}
\end{table}

\begin{figure}
\begin{center}
$\begin{array}{c}
\mbox{(a)} \\
\includegraphics[scale=.55]{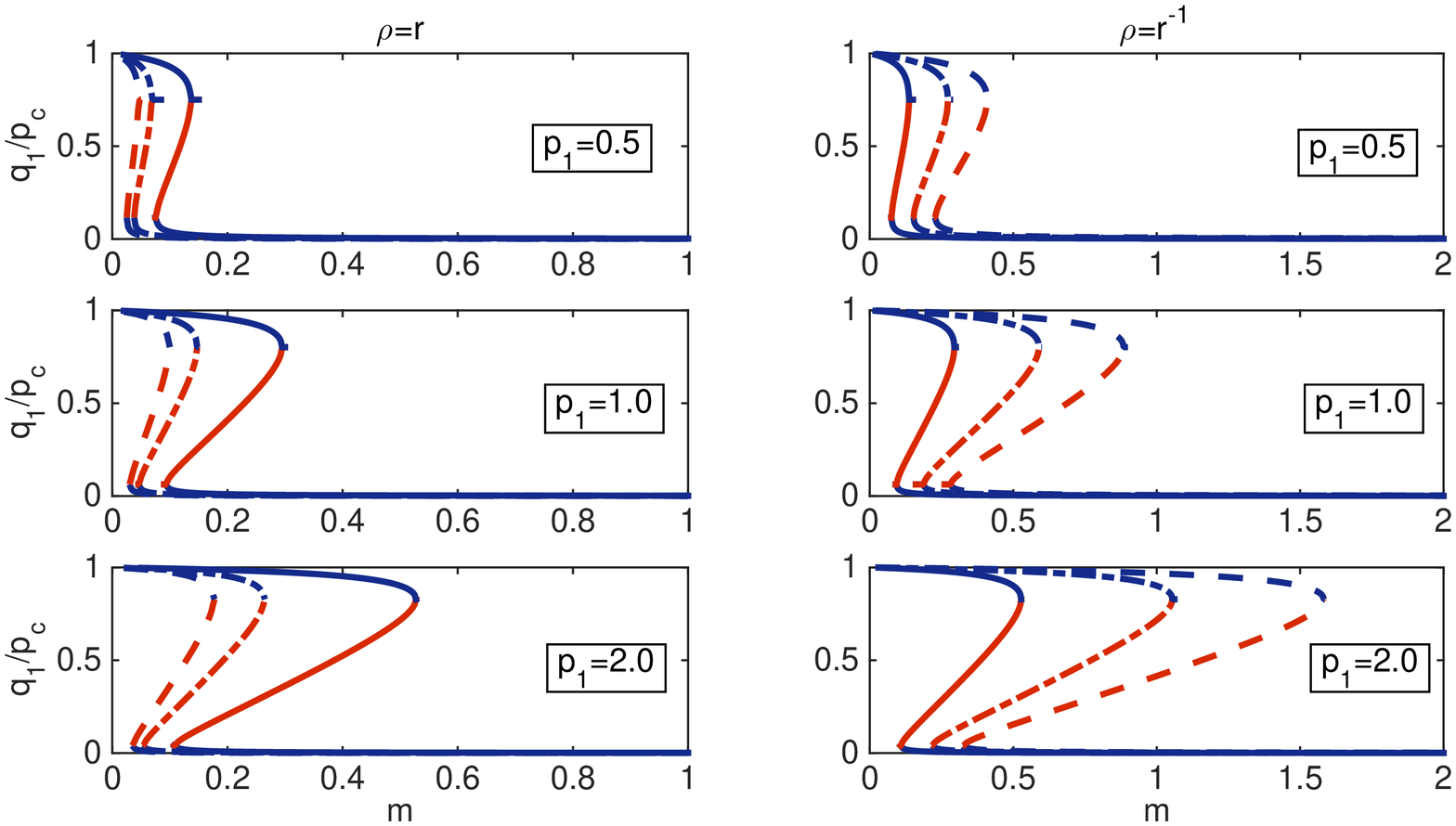} \\
\mbox{(b)} \\
\includegraphics[scale=0.370]{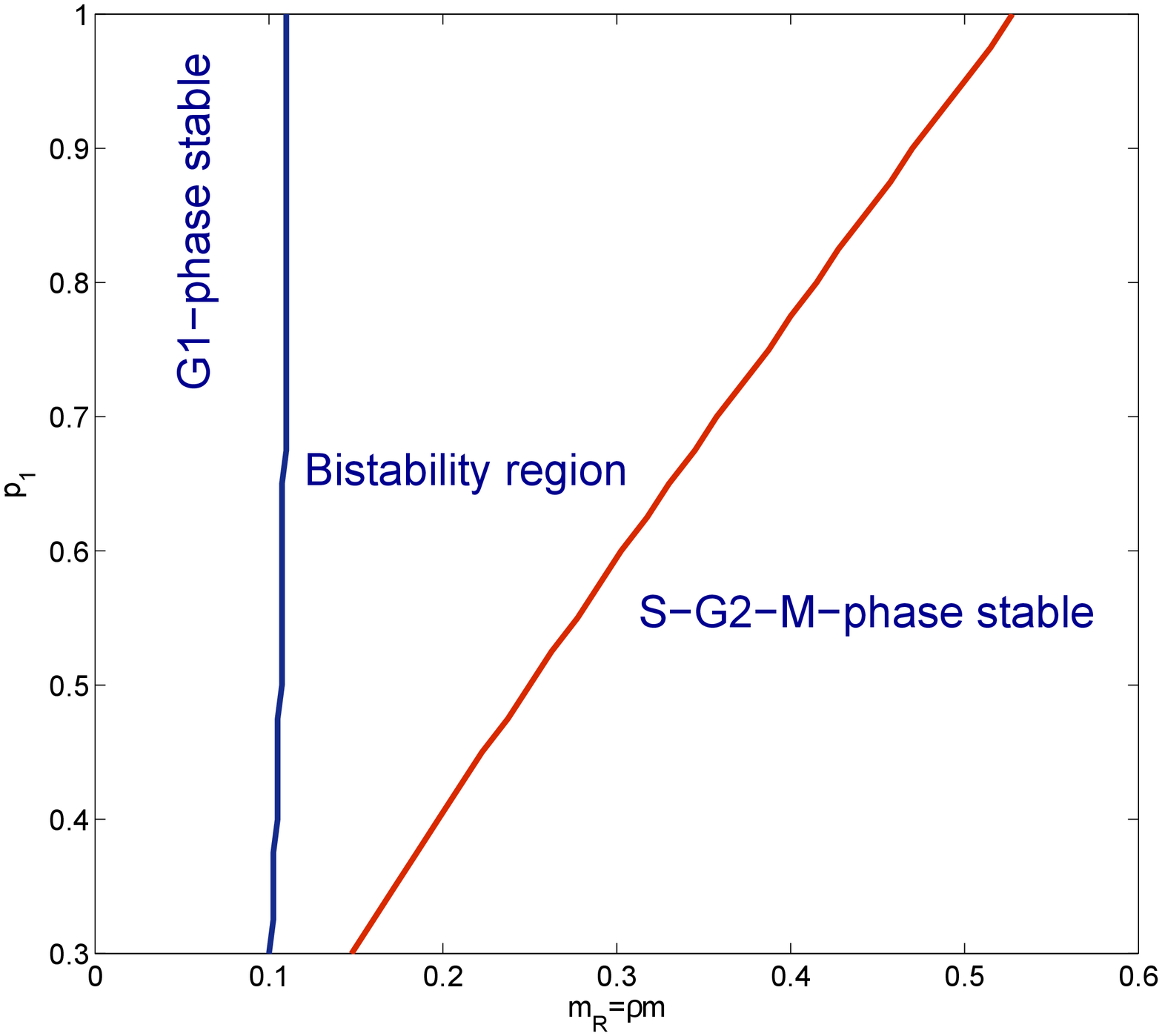}
\end{array}$
\caption{(a) Bifurcation analysis for the SCQSS approximation of the stochastic bistable enzyme-catalysed system Eqs.
(\ref{eq:hjqss-x1})-(\ref{eq:hjqss-p4}). The panels on the top plot (a) shows
the bifurcation diagrams for different values of the parameters $p_1$, $p_{c}=p_{1}$ and $\rho=\frac{p_3p_{e_3}}{p_4p_{e_4}}$. If $e_0$ and $s_0$ are
random Poisson variables with parameter $S$ and $E$, respectively, then $\rho=\frac{p_3^2}{p_4^2}$ (see Eq. \ref{eq:psenzyme}). In these panels solid
lines correspond to $r=1$,
dot-dashed lines to $r=2$, and dashed lines to $r=3$. The bottom plot (b) shows the bi-stability boundaries in $p_1-m_R$ parameter space. The region
between the boundaries corresponds to the bistable region of the stochastic Tyson \& Novak system according to the SCQSS approximation.
}\label{fig:bifanTN}
\end{center}
\end{figure}

Upon rescaling of the variables (Table \ref{tab:dless-tn}) and the Hamiltonian (Eq. (\ref{eq:scaledham})), the action functional reads:

\begin{eqnarray}\label{eq:scaledactionintegral}
\nonumber S(p,q)=&&s_0\int_0^\tau\left(-H_{\kappa}(p,q)-\sum_{slow}q_i\frac{dp_i}{ds}-\sum_{fast}q_j\epsilon\frac{dp_j}{ds}\right)ds \\
&& + \sum_i^nS_{0,i}(p_i)
\end{eqnarray}

\noindent It is straightforward to check that in SCQSSA conditions $H_{\kappa}(p,q)=0$. Furthermore, since $p_1=p_2=$const. and $p_7=1$, and
$\epsilon\dot{p_j}\simeq 0$ for the fast generalised coordinates, the SCQSS approximation of the action Eq. (\ref{eq:scaledactionintegral}),
$S_{QSS}$, reduces to:

\begin{equation}\label{eq:qssactionintegral}
S_{QSS}(p)=\sum_{i=1}^nS_{0,i}(p_i)
\end{equation}

\noindent where, as per the SCQSSA, $p_5$ and $p_6$ are determined by Eqs. (\ref{eq:hjqss-p3}) and (\ref{eq:hjqss-p4}), respectively, $p_7=1$, which
implies $S_{0,7}(p_7)=0$, and $p_1=p_2$, $p_3$ and $p_4$ are constants that remain to be determined. In order to do so, we resort to the method
developed in reference \cite{alarcon2014}. The quasi-steady state characteristic function, $G_{QSS}(p,\tau)$ is given by:

\begin{equation}\label{eq:qssgenerfunc} 
G_{QSS}(p,\tau)=e^{\left(-\sum_{i=1}^6S_{0,i}(p_i)\right)}=\prod_{i=1}^6G_{0,i}(p_i)
\end{equation}

\noindent where $G_{0,i}(p_i)=e^{-S_{0,i}(p_i)}$ is the generating function of the probability distribution for the initial condition of species $X_i$
$i=1,\dots,6$. In \cite{alarcon2014}, we have shown that, applying a Laplace-type asymptotic method \cite{murray1984,ablowitz2003} to the integrals

\begin{eqnarray}\label{eq:cauchyintegrals}
\nonumber && P(X_1(\tau=0)=s_0)=\frac{1}{2\pi i}\oint_C\frac{G_{0,1}(p_1)}{p_1^{s_0+1}}dp_1=\frac{1}{2\pi i}\oint_C\frac{e^{-(S_{0,1}(p_1)+s_0\log
p_1)}}{p_1}dp_1,\\
&& P(X_i(\tau=0)=e_i)=\frac{1}{2\pi i}\oint_C\frac{e^{-(S_{0,i}(p_i)+e_0\log p_i)}}{p_i}dp_i\mbox{ with }i=3,4,
\end{eqnarray}

\noindent where, $p_1=p_2$, $p_3$ and $p_4$ can be given as functions of $s_0$ and $e_i,\,i=3,4$, i.e. the initial numbers of Cdh1 molecules and
Cdh1-inactivating and Cdh1-activating enzymes, respectively:

\begin{eqnarray}\label{eq:laplaceapprox}
\nonumber && -p_1\frac{dS_{0,1}}{dp_1}=s_0 \\
&& -p_i\frac{dS_{0,i}}{dp_i}=e_i\mbox{ for } i=3,4
\end{eqnarray}

\noindent $P(X_1(\tau=0)=s_0)$, $P(X_3(\tau=0)=e_3)$ and $P(X_4(\tau=0)=e_4)$ are the probabilities that $X_1$ initially takes the value
$X_1(\tau=0)=s_0$ and that $X_3$ and $X_4$ have initial values $X_3(\tau=0)=e_3$ and $X_4(\tau=0)=e_4$. These probabilities can be interpreted to correspond to
variability in the abundance of these enzymes within a population of cells. A particularly simple case results from assuming that
$P(X_1(\tau=0)=s_0)$, $P(X_3(\tau=0)=e_3)$ and $P(X_4(\tau=0)=e_0)$ are Poisson distributions with parameter $S$ and $E$, respectively. In this case
\cite{alarcon2014}:

\begin{eqnarray}\label{eq:psenzyme}
\nonumber && p_1=\frac{s_0}{S} \\
\nonumber && p_3=\frac{e_3}{E} \\
&& p_4=\frac{e_4}{E}
\end{eqnarray}

\noindent Note that, in the particular case in which the total numbers of Cdh1 and enzyme molecules are random Poisson variables, we have that
$p_1=p_c$, $p_3=p_{e_3}$, and $p_4=p_{e_4}$. 

\subsection{Bifurcation analysis}

\begin{figure}
\begin{center}
\includegraphics[scale=0.4]{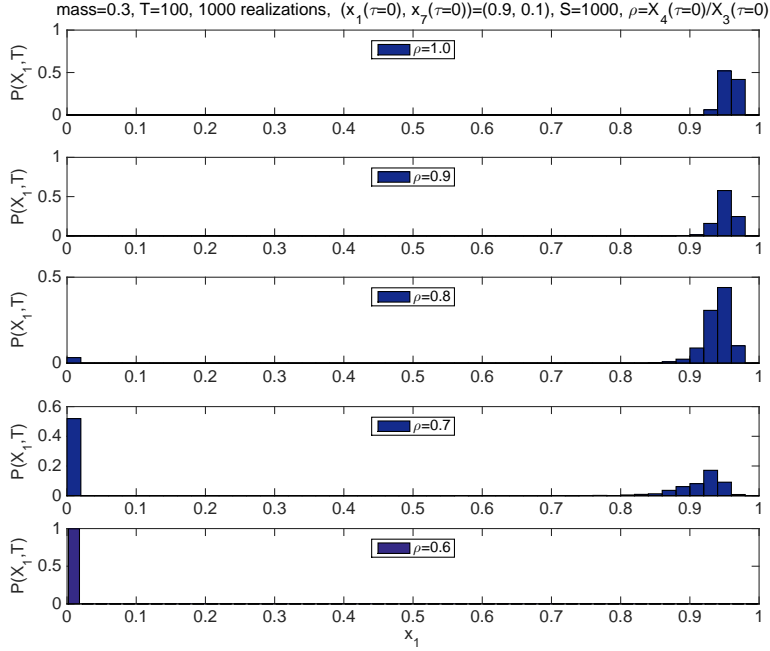}
\caption{Simulation results for the stochastic bistable enzyme-catalysed system Table \ref{tab:smm-bistable}. We have plotted the probability
$P(x_1,T)=\mbox{Prob}(x_1(\tau=T))$ where $x_1=X_1/S$ and $T=100$ for different values of $\rho$.  The initial number of Cdh1-inactivating and Cdh1-activating enzymes are fixed according to
$X_3(t=0)=\frac{e_0}{\rho}$ and $X_4(t=0)=e_0$, respectively. $m=0.3$. We aim to check our predictions regarding the effect of the ratio $\rho=\frac{p_3^2}{p_4^2}=\frac{e_3^2}{e_4^2}$ on the stability properties of the system. According to our results shown in Fig. \ref{fig:bifanTN}, decreasing the ratio between the number of Cdh1-inactivating ($e_4$) and Cdh1-activating ($e_3$) enzymes, the system should be driven away from bistability and into the stable G$_1$-phase regime (see Fig. \ref{fig:bifanTN}(b)). The remaining parameter values are inferred from those given by Tyson \& Novak
\cite{tyson2001} as shown in Tables \ref{tab:dless-tn} and \ref{tab:par-val}. We see that when varying $\rho$, the system switches from a state of high $x_1$ ($\rho\geq 0.9$) ro a state of low $x_1$ ($\rho\leq 0.6$), whereas at the intermediate levels of (e.g. $\rho=0.7$ and $\rho=0.8$) the system is in a bistable state. We take $p_1=p_c=1$ in all the simulations shown in this figure. Average is
performed over 1000 realisations.}\label{fig:ratios}
\end{center}
\end{figure}

Fig. \ref{fig:bifanTN} shows results regarding the bifurcation behaviour of the SCQSS approximation of the stochastic bistable enzyme-catalysed system
Eqs. (\ref{eq:hjqss-x1})-(\ref{eq:hjqss-p4}). In particular we are interested in a comparison between the bistable behaviour of the mean-field model,
corresponding to taking $p_i=1$ for all $i$, and that of the SCQSS approximation with $p_1$, $p_3$ and $p_4$ given by Eq. (\ref{eq:laplaceapprox}).
i.e. they are determined as functions of $s_0$ and $e_0$. 

We have shown that both the ratio of $p_3$ and $p_4$, $\rho=\frac{p_3p_{e_3}}{p_4p_{e_4}}=\frac{p_3^2}{p_4^2}=\frac{e_3^2}{e_4^2}$, and $p_1$ alter the bistable behaviour of the system beyond the
predictions of the mean-field model. In particular, we observe that decreasing the value of $\rho$ extends the region of stability of the G$_1$-fixed
point, i.e. the fixed point corresponding to the steady-state value of $q_1$, such that $q_1\sim 1$. By contrast, when $\rho$ is increased
the
stability region of the G$_1$-fixed point shrinks. Intuitively, given the relation between $p_3$ and $p_4$ and the number of Cdh1-inactivating and
Cdh1-activating enzyme, this result is straightforward to interpret: decreasing the number of Cdh1-inactivating enzyme demands a larger value of $m$
in order to de-stabilise the G$_1$-fixed point. This is fully confirmed by direct simulation using Gillespie stochastic simulation algorithm
\cite{gillespie1976}. Fig. \ref{fig:ratios} shows simulation results in which we compute the probability $P(x_1,T)=\mbox{Prob}(x_1(\tau=T))$ 
for different values of $\rho\leq 1$. $T$ has been chosen so that the system has reached steady state conditions. We observe, that for $\rho=1$ and $m=0.3$, the system evolves towards the
$q_1\ll 1$-fixed point (i.e. the S-G$_2$-M fixed point). As $\rho$ decreases, i.e. there is more Cdh1-inactivating enzyme than
Cdh1-activating enzyme, the system enters the bistable regime. If $\rho$ reaches low-enough values (depending upon the initial condition), we may even
observe an exchange of stability, i.e. the system evolves towards the $q_1\sim 1$-fixed point.   

\begin{figure}
\begin{center}
\includegraphics[scale=0.4]{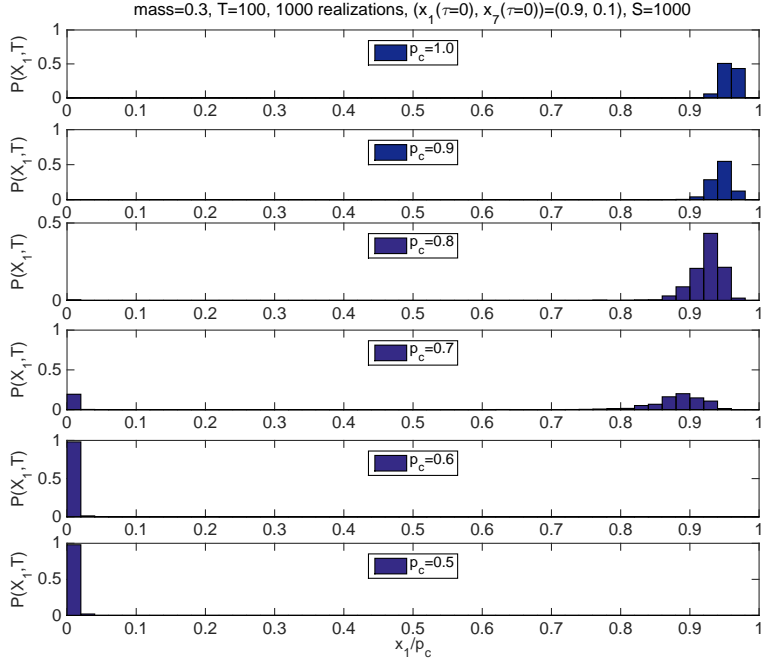}
\caption{Simulation results for the stochastic bistable enzyme-catalysed system Table \ref{tab:smm-bistable}. We have plotted the probability
$P(x_1,T)=\mbox{Prob}(x_1(\tau=T))$ where $x_1=X_1/S$ and $T=100$ with different initial conditions and different values of $p_c$. Average is
performed over 1000 realisations. $m=0.3$ and $X_3(t=0)=e_0$ and $X_4(t=0)=e_0$. The remaining
parameter values are inferred from those given by Tyson \& Novak \cite{tyson2001} as shown in Tables \ref{tab:dless-tn} and
\ref{tab:par-val}. We see that when varying $p_c$, the system switches from a state of high $x_1$ ($p_c\geq 0.8$) ro a state of low $x_1$ ($p_c\leq 0.6$), whereas at the intermediate levels of $p_c=0.7$ the system is in a bistable state.}\label{fig:p1}
\end{center}
\end{figure}

Regarding the dependence on $p_1$, we have checked the predictions of the SCQSS approximation by means of simulations with different values of $s_0$.
Figure \ref{fig:bifanTN} shows the bi-stability region of system Eqs. (\ref{eq:hjqss-x1})-(\ref{eq:hjqss-p4}) in $p_1-m_R$-space, where $m_R=\rho m$.
For a fixed value of $m_R$, there is a threshold value for $p_1$ below which the system stops being bistable to become entrapped into the the
S-G$_2$-M fixed point (i.e. $q_1\ll 1$). In order to validate this prediction, we have conducted stochastic simulations for different values of $s_0$.
Figure \ref{fig:p1} shows simulation results for $P(x_1,T)=\mbox{Prob}(x_1(\tau=T))$. We observe that for small values of $s_0$, the system is locked 
into the the S-G$_2$-M fixed point, as predicted by the SCQSS approximation. As $s_0$ increases, the system enters a fluctuation-dominated bistable 
regime where, as the system goes through the bifurcation point, the system undergoes bistable behaviour. This behaviour is typical in a system 
undergoing a phase transition, where fluctuations unboundedly increase \cite{goldenfeld1992}. Finally, as $s_0$ continues to increase, the system 
becomes trapped into G$_1$-fixed point (see Figure \ref{fig:p1}). These results fully reproduce the behaviour predicted by our SCQSSA stability
analysis.

\begin{figure}
\begin{center}
$\begin{array}{cc}
\mbox{(a)} & \mbox{(b)} \\
\includegraphics[scale=0.3]{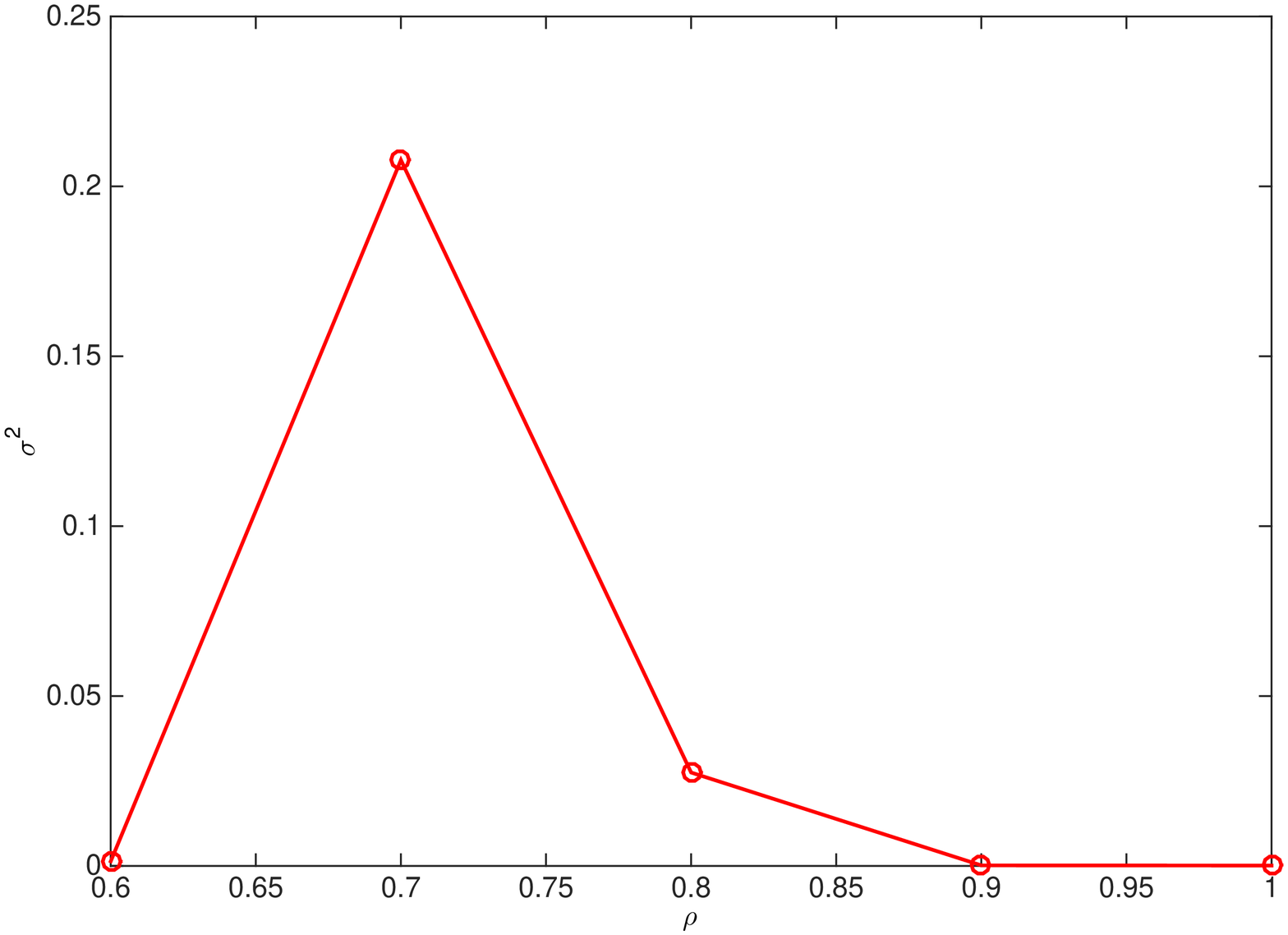} & \includegraphics[scale=0.33]{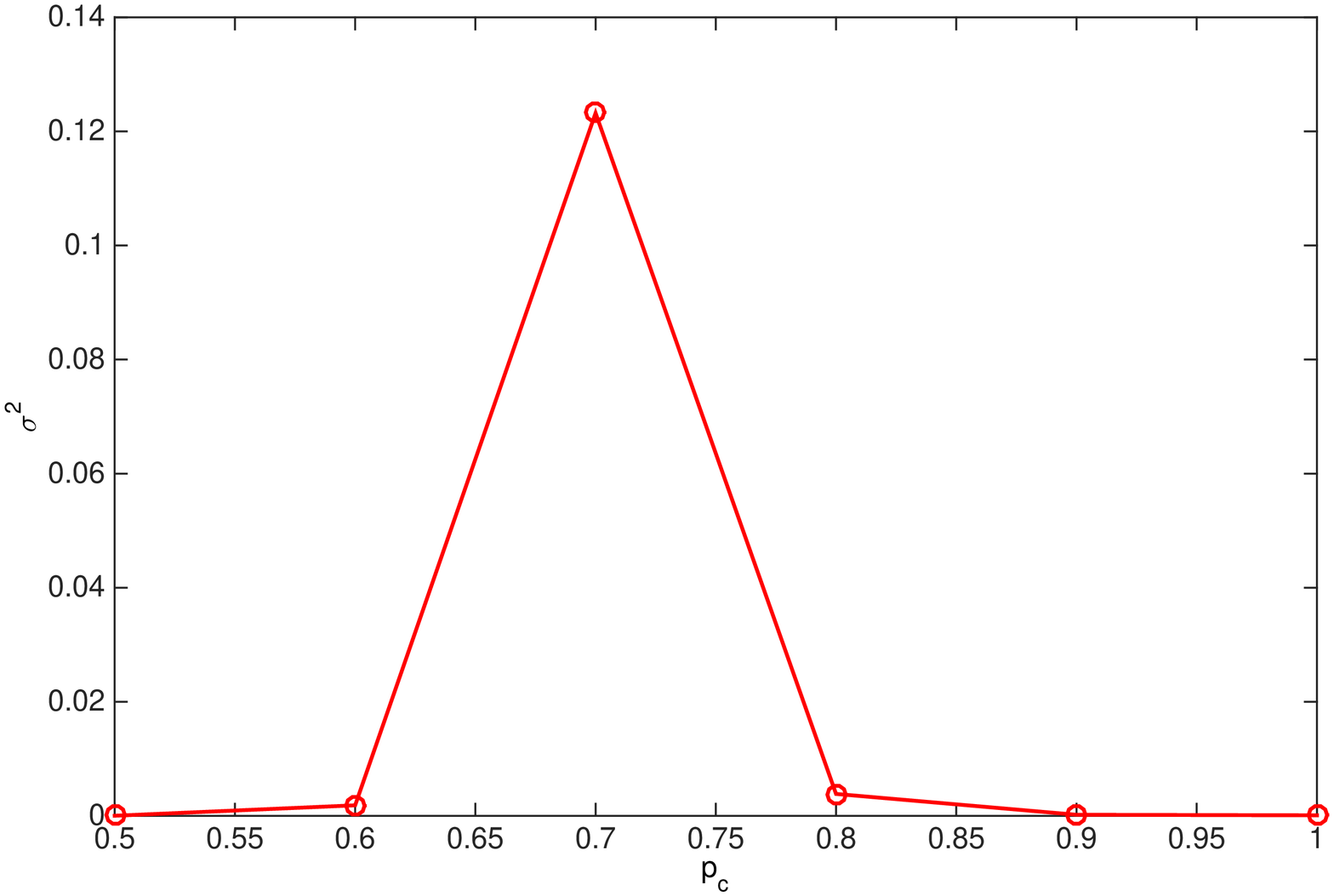}
\end{array}$
\caption{Plots showing the variance $\sigma^2=\langle(x_1-\langle x_1\rangle)^2\rangle$ where $x_1=X_1(\tau=T)/S$ 
associated to the simulation results shown in Fig. \ref{fig:ratios} (panel (a)) and in Fig. \ref{fig:p1} (panel (b)). These plots show how 
$\sigma^2$ changes as the control parameter ($\rho$, for the simulations associated to plot (a), and $p_c$  for the simulations shown in plot 
(b)). The maximum of $\sigma^2$ as a function of the control parameter helps us to quantitatively determine the corresponding critical 
value \cite{goldenfeld1992}.}\label{fig:sigma2enzyme} 
\end{center}
\end{figure}

The aforementioned behaviour regarding unbounded increase of fluctuations close to a bifurcation \cite{goldenfeld1992} is used to locate the critical 
value of the associated control parameter, i.e. $\rho$ and $p_c$ for the simulations shown in Figs. \ref{fig:ratios} and \ref{fig:p1}, respectively. 
This property allows us to do a quantitative comparison between the simulations and asymptotic analysis. To this end, we plot how the variance, 
$\sigma^2=\langle(x_1-\langle x_1\rangle)^2\rangle$ where $x_1=X_1(\tau=T)/S$, changes as the corresponding control parameter varies. Regarding the 
results shown in Fig. \ref{fig:sigma2enzyme}(a) (associated to the simulations shown in Fig. \ref{fig:ratios}), we observe that the critical value of 
the control parameter $\rho$, $\rho_B$, is approximately $\rho_B\simeq 0.7$, which, taking into account that $m=0.3$, implies that the critical value 
of the renormalized mass, $m_R=\rho m$, $m_B=\rho_B m\simeq 0.21$. Our asymptotic analysis predicts that $m_B=0.11$ (see Fig. \ref{fig:bifanTN}(b) 
with $p_c=1$). The results shown in in Fig. \ref{fig:sigma2enzyme}(b) (corresponding to the simulations shown in Fig. \ref{fig:p1}), the critical 
value of $p_c$, $p_B$, is approximately $p_B\simeq 0.7$. The prediction of our asymptotic analysis (see Fig. \ref{fig:bifanTN}(b) 
with $\rho=1$) is $p_B=0.6$. 

\section{Auto-activation gene regulatory circuit}\label{sec:gene}

We now proceed to analyse the effects of intrinsic noise in a model of a bistable self-activation gene regulatory circuit 
\cite{assaf2011,frigola2012,weber2013}
in the context of the quasi-steady regime. Many instances of genetic switches, i.e. bistable gene regulatory circuits, have been identified
\cite{gardner1999,ozbudak2004,yao2008,lee2010,yao2014}. Most of them are characterised by the presence of a positive feed-back in which one of the
molecular species involved in the system up-regulates its own production. All of these systems exhibit bi-stability and hysteresis, i.e. a form of
memory associated to bistable systems, and some of them are thought to exist in regimes where stochastic switching is frequent
\cite{acar2005,lee2010}. Noise effects on this kind of system has been extensively analysed and found to have both constructive and deleterious
effects. For example, Frigola et al. \cite{frigola2012} have found that noise stabilises the inactive (OFF) steady-state of a model of a bistable
self-activation gene 
regulatory circuit by extending its stability region. In this Section, we analyse the effects of noise specifically associated to the quasi-steady
state regime in the large-deviations (large number of molecules) limit. 

\begin{figure}
\begin{center}
\includegraphics[scale=0.4]{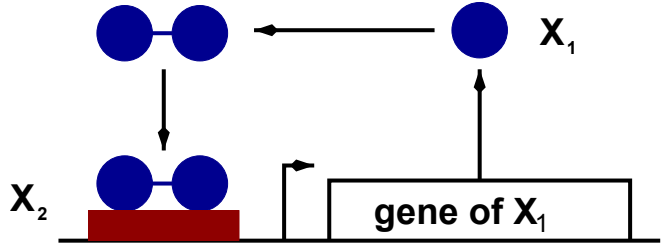} 
\caption{Schematic representation of the self-activating gene regulatory circuit. The gene product $X_1$ is its own transcription factor which, upon
dimerisation, binds the promoter region of the gene thus triggering gene transcription. The transition rates corresponding to this gene regulatory
circuit are given in Table \ref{tab:smm-gene}. For simplicity, we use an effective model in which the formation of the dimer and binding to the promoter region is taken into account in a single reaction, and the resulting number of promoter sites bound by two transcription factors is denoted $X_2$.}\label{fig:selactscheme}
\end{center}
\end{figure}  

\begin{table}[htb]
\begin{center}
\begin{tabular}{ll}
Variable & Description\\\hline
$X_1$ & Number of transcription factor molecules \\
$X_2$ & Number of bound promoter sites  in the gene promoter region \\
$X_3$ & Number of unoccupied (unbound) binding sites in the gene promoter region\\\hline
\end{tabular}
\begin{tabular}{lll}
Transition rate & r & Event \\\hline
$W_1(x)=\hat{R}+k_{1}X_2$ & $r_1=(1,0,0)$ & Synthesis of the transcription factor\\
$W_2(x)=k_{2}X_1$ & $r_2=(-1,0,0)$ & Degradation of the transcription factor\\
$W_3(x)=k_{3}X_1(X_1-1)X_3$ & $r_3=(-2,+1,-1)$ & Dimer binding to the gene promoter region \\
$W_4(x)=k_{4}X_2$ & $r_4=(+2,-1,+1)$ & Unbinding from the gene promoter region
\end{tabular} 
\end{center}
\caption{Random variables and transition rates associated to the stochastic dynamics of an auto-activation gene regulatory circuit
\cite{frigola2012,weber2013}. $X_2$ corresponds to the number of transcription-factor dimer/promoter binding site trimers. See Fig.
\ref{fig:selactscheme} for an schematic representation.}\label{tab:smm-gene}
\end{table}

We study the stochastic system of the simple self-activating gene regulatory circuit schematically represented in Fig. \ref{fig:selactscheme}. In
this circuit the gene product binds to form dimers which then act as its own transcription factor by binding to the promoter region of the gene. The
rate-limiting factor is therefore the number of available binding sites within the promoter of the gene. For simplicity, our stochastic model
associated to the rates shown in Table \ref{tab:smm-gene} does not explicitly account for dimer formation. We will assume that this process is very
fast so it can be subsumed under the formation of transcription-factor dimer/promoter binding site trimers (reaction 3, Table \ref{tab:smm-gene}).
Furthermore, it is important to note that our stochastic dynamics exhibits a conservation law: $X_2+X_3=e_0$ at all time. This conservation law
expresses the fact that the total number of binding sites, $e_0$, is constant. 

In order to proceed with our analysis of the stochastic model of self-activated gene regulation (see Table \ref{tab:smm-gene} and Fig.
\ref{fig:selactscheme}), we apply the general methodology associated to our SCQSS approximation. Following the general procedure explained in the
previous sections, we start by deriving the stochastic Hamiltonian associated to the process defined by the transition rates shown in Table
\ref{tab:smm-gene} (see Section \ref{sec:theory}):

\begin{equation}\label{eq:selfactham}
H(p,Q)=(p_{1}-1)(\hat{R}+k_{1}Q_{2}p_{2})+k_{2}(1-p_{1})Q_{1}+k_{3}(p_{2}-p_{1}^{2}p_{3})Q_{1}^{2}Q_{3}+k_{4}(p_{1}^{2}p_{3}-p_{2})Q_{2}, 
\end{equation}

\noindent which, according to our theory (see Section \ref{sec:theory}), gives raise to the re-scaled Hamiltonian, $H_{\kappa}(p,q)$, defined by,

\begin{equation}\label{eq:rescselfactham}
H_{\kappa}(p,q)=(p_{1}-1)(R+\kappa_{1}q_{2}p_{2})+\kappa_{2}(1-p_{1})q_{1}+(p_{2}-p_{1}^{2}p_{3})q_{1}^{2}q_{3}+\kappa_{4}(p_{1}^{2}p_{3}-p_{2}
)q_{2}, 
\end{equation}

\noindent where $H(p,Q)=k_3ES^2H_{\kappa}(p,q)$ and the re-scaled variables, $q_i$, and re-scaled rate constants, $\kappa_j$, are defined in Table
\ref{tab:dless-selfact}.

\begin{table}[htb]
\begin{center}
\begin{tabular}{lll}
Rescaled variables & \vline \mbox{ } & Dimensionless parameters\\\hline 
$\tau=k_3ESt$ & \vline \mbox{ }  & $\epsilon=E/S$, $R=\hat{R}/(k_3ES^2)$\\
$q_1=Q_1/S$ & \vline \mbox{ }  & $\kappa_1=k_1/(k_3S^2)$ \\
$q_2=Q_2/E$ & \vline \mbox{ }  & $\kappa_2=k_2/(k_3ES)$\\
$q_3=Q_3/E$ & \vline \mbox{ }  & $\kappa_4=k_4/(k_3S^2)$
\end{tabular} 
\end{center}
\caption{Dimensionless variables used in Eqs. (\ref{eq:kappaham}). $s_0$ is a characteristic scale associated to the average number of molecules of
transcription factor, $X_1$, and $E$ is the average number of binding sites in the promoter of the self-activating gene. We further assume that $S\gg
E$.}\label{tab:dless-selfact}
\end{table}

The re-scaled Hamilton equations are thus given by:

\begin{eqnarray}
\label{eq:res-sa-q1} \frac{dq_{1}}{d\tau}&=&R+\kappa_{1}q_{2}p_{2}-\kappa_{2}q_{1}-2q_{1}^{2}q_{3}p_{1}p_{3}+2\kappa_{4}p_{1}p_{3}q_{2}
\\
\label{eq:res-sa-q2} \epsilon\frac{dq_{2}}{d\tau}&=&(p_{1}-1)\kappa_{1}q_{2}+q_{1}^{2}q_{3}-\kappa_{4}q_{2}
\\
\label{eq:res-sa-q3} \epsilon\frac{dq_{3}}{d\tau}&=&-q_{1}^{2}q_{3}p_{1}^{2}+\kappa_{4}p_{1}^{2}q_{2}
\\
\label{eq:res-sa-p1} \frac{dp_{1}}{d\tau}&=&\kappa_{2}(p_{1}-1)-2q_{1}q_{3}(p_{2}-p_{1}^{2}p_{3})
\\
\label{eq:res-sa-p2} \epsilon\frac{dp_{2}}{d\tau}&=&\kappa_{1}(1-p_{1})p_{2}-\kappa_{4}(p_{1}^{2}p_{3}-p_{2})
\\
\label{eq:res-sa-p3} \epsilon\frac{dp_{3}}{d\tau}&=&q_{1}^{2}(p_{1}^{2}p_{3}-p_{2})
\end{eqnarray}

\noindent From these equations, we observe that for $q_2(\tau)+q_3(\tau)=p$, where $p=e_0/E$, to hold we must have that $p_1(\tau)=1$ for
all $\tau$. Imposing this
condition on Eq. (\ref{eq:res-sa-p1}) implies that $p_2(\tau)=p_3(\tau)$, which, in turn, together with Eqs. (\ref{eq:res-sa-p2}) and
(\ref{eq:res-sa-p3}), imply that $p_2=p_3=$const. Finally, applying the QSS approximation to remaining equations, Eqs.
(\ref{eq:res-sa-q1})-(\ref{eq:res-sa-q3}), we obtain:

\begin{eqnarray}
\label{eq:qss-sa-q1}\frac{dq_{1}}{d\tau}=R+\kappa_{1}pp_{2}\frac{q_{1}^{2}}{\kappa_{4}+q_{1}^{2}}-\kappa_{2}q_{1},\\
\label{eq:qss-sa-q2}q_{2}=p-q_3=p\frac{q_{1}^{2}}{\kappa_{4}+q_{1}^{2}}.
\end{eqnarray}

\noindent As for the bistable enzyme-catalysed system, the parameter values are determined by matching the mean-field limit of our stochastic model,
which is obtained by setting $p_i=1$ for all $i$ \cite{elgart2004} and $p=1$ (i.e. the number of binding sites exactly equal to its average), to the
mean-field system proposed by Frigola et al.
\cite{frigola2012}. The mapping of our parameters to those of reference \cite{frigola2012} and their associated values are given in Table
\ref{tab:par-val-selfact}.

\begin{table}[htb]
\begin{center}
\begin{tabular}{llll}
Rescaled parameter & \vline Parameter & \vline \mbox{Units} & \vline Reference\\\hline 
$\kappa_1=\frac{a}{k_{deg}\sqrt{K_d}}$ & \vline $K_d=10$ & \vline \mbox{nM} & \vline   \cite{frigola2012} \\
$\kappa_2=1$ & \vline $k_{deg}=2$ & \vline \mbox{min}$^{-1}$ & \vline   \cite{frigola2012} \\
$\kappa_4=1$ & \vline $r=0.4$ & \vline \mbox{nM} $\cdot$ \mbox{min}$^{-1}$ & \vline   \cite{frigola2012} \\
$R=\frac{r}{k_{deg}\sqrt{K_d}}$ & \vline $S=1.0$ & \vline \mbox{} & \vline   \mbox{--} \\
$k_3ES=k_{deg}$ & \vline $E=0.1$ & \vline \mbox{} & \vline   \mbox{--} \\
\end{tabular} 
\end{center}
\caption{Parameter values used in simulations of the stochastic self-activation gene regulatory circuit.}\label{tab:par-val-selfact}
\end{table}

Finally, according to the theory developed in Section \ref{sec:theory}, $p_2$ is determined in terms of the total number of binding sites within the
gene promoter, $e_0$:

\begin{equation}\label{eq:selfact-p2}
-p_2\frac{dS_0}{dp_2}=e_0, 
\end{equation}

\noindent where $S_0(p)=\ln(G_0(p))$ and $G_0(p)$ is the generating function associated to the probability distribution of $e_0$, $P(e_0)$. This
probability distribution can be interpreted as corresponding to the distribution over a cell population of the number of binding sites in the promoter
of gene $x_1$. For example, if $P(e_0)$ is a Poisson distribution the Eq. (\ref{eq:selfact-p2}) reads \cite{alarcon2014}

\begin{equation}\label{eq:selfact-p2-poisson}
p_2=\frac{e_0}{E}, 
\end{equation}

\noindent where $E\equiv\langle e_0\rangle$, i.e. the average of $e_0$ over a population of cells. Therefore, according to this analysis, we have that
$p=p_2$, provided that $P(e_0)$ is a Poisson distribution with parameter $E$.

\subsection{Bifurcation analysis}

\begin{figure}
\begin{center}
\includegraphics[scale=0.6]{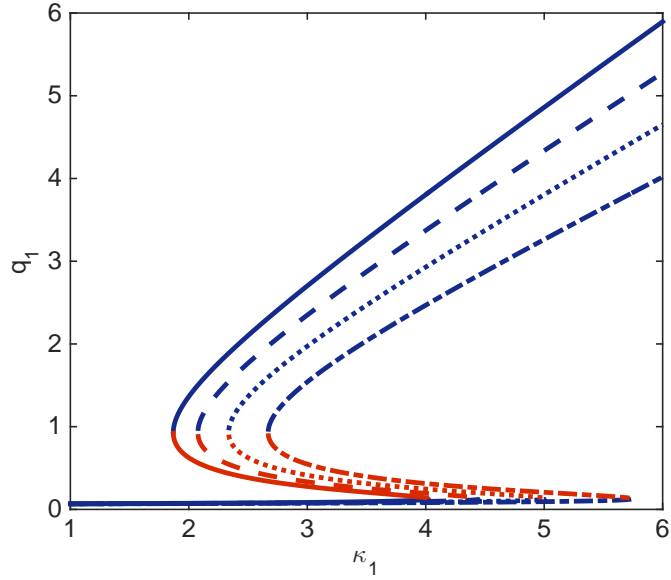} 
\caption{Bifurcation analysis for the SCQSS approximation of the stochastic auto-activation gene regulatory circuit Eqs.
(\ref{eq:qss-sa-q1})-(\ref{eq:qss-sa-q2}). This figure shows the bifurcation diagram for different values of the parameters of $p_2$. In these panels
solid lines correspond to $p_2^2=1$, dashed lines to $p_2^2=0.9$, dotted lines to $p_2^2=0.8$, and dash-dotted lines to $p_2^2=0.7$ (recall that
$p_2=p$). Parameter values as
given
in Table \ref{tab:par-val-selfact}.}\label{fig:bifdiagselfact}
\end{center}
\end{figure}

Fig. \ref{fig:bifdiagselfact} shows results regarding how the bifurcation diagram varies as we change $p p_2=p_{2}^{2}$, which, we recall, is determined by the
(probability distribution of the) total number of binding sites within the gene promoter. Inspection of Eq. (\ref{eq:qss-sa-q1}) shows that $p_2$ has
the effect of renormalising the self-activation rate $\kappa_1$. If $p_2^{2}<1$ then the rate of gene self-activation is effectively reduce and,
consequently the stability region of the inactive steady-state, $q_1\sim 0$, is extended. That is, we need to go to larger values of $\kappa_4$ to
enter
the region where the active steady-state, $q_1> 1$, becomes stable (see Fig. \ref{fig:bifdiagselfact}). On the contrary, $p_2^{2}>1$ has the effect of
extending the stability region of the active steady-state, $q_1> 1$. 

\begin{figure}
\begin{center}
\includegraphics[scale=0.5]{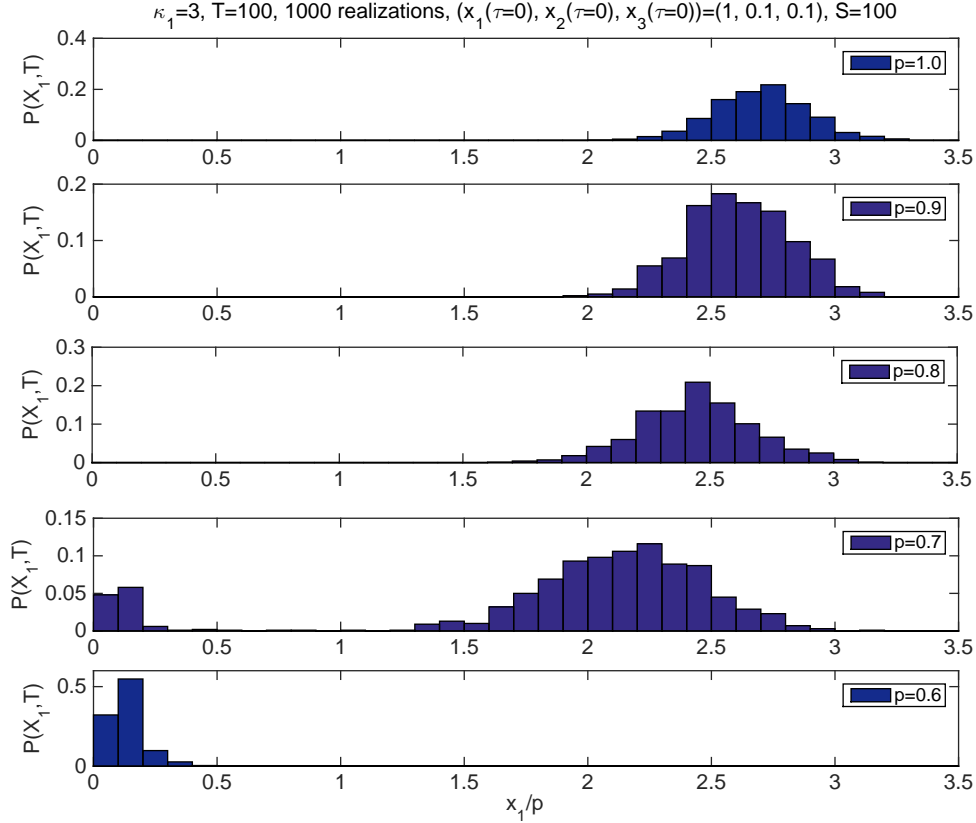}  
\caption{Simulation results for the stochastic gene regulatory circuit of self-activation (Table \ref{tab:smm-gene}). We have plotted the probability
$P(x_1,T)=\mbox{Prob}(x_1(\tau=T))$ where $x_1=X_1/S$ and $T=100$ as the number of binding sites in the gene promoter, given by $X_3(t=0)=pE$. Average is performed over 1000 realisations. Parameter values are inferred from those given by Frigola et al. \cite{frigola2012}
as shown in Tables \ref{tab:dless-selfact} and \ref{tab:par-val-selfact}. We see the emergence of bistability at $p=0.7$, whereas for smaller(larger) values of $p$, the system will be in the stable steady state corresponding to low(high) number of transcription factor molecules.}\label{fig:histselfact}
\end{center}
\end{figure}

In order to verify the predictions of our bifurcation analysis (Fig. \ref{fig:bifdiagselfact}), we consider Eqs. (\ref{eq:selfact-p2}) and
(\ref{eq:selfact-p2-poisson}), which relate the momentum variable $p_2$ to the number of binding sites within the gene promoter. If we assume that the
latter is distributed according to a Poisson distribution, then Eq. (\ref{eq:selfact-p2-poisson}) holds and $p_2=p= e_0/E$. Under these
conditions, our bifurcation analysis predicts that the probability distribution of $X_1$, i.e. the random variable associated to the generalised
coordinate $q_1$, should change, as $e_0$ decreases, from being uni-modal with a single maximum about the ON value of $X_1$ (or, when scaled with
$s_0$, $q_1$) to exhibiting bi-modality, as the system approaches the saddle-node bifurcation which annihilates the ON state as it collides with the
saddle point, with two peaks about the ON and OFF states. If $e_0$ is further reduced the system will be driven passed this saddle-node bifurcation,
the 
probability distribution becomes uni-modal but, unlike its large $e_0$ counterpart, its peak is about the OFF $q_1$-steady-state. We have verified
this
prediction by running simulations using the SSA. The results, which agree with our prediction, are shown in Fig. \ref{fig:histselfact}.

\begin{figure}
\begin{center}
$\begin{array}{cc}
\mbox{(a)} & \mbox{(b)}\\
\includegraphics[scale=0.35]{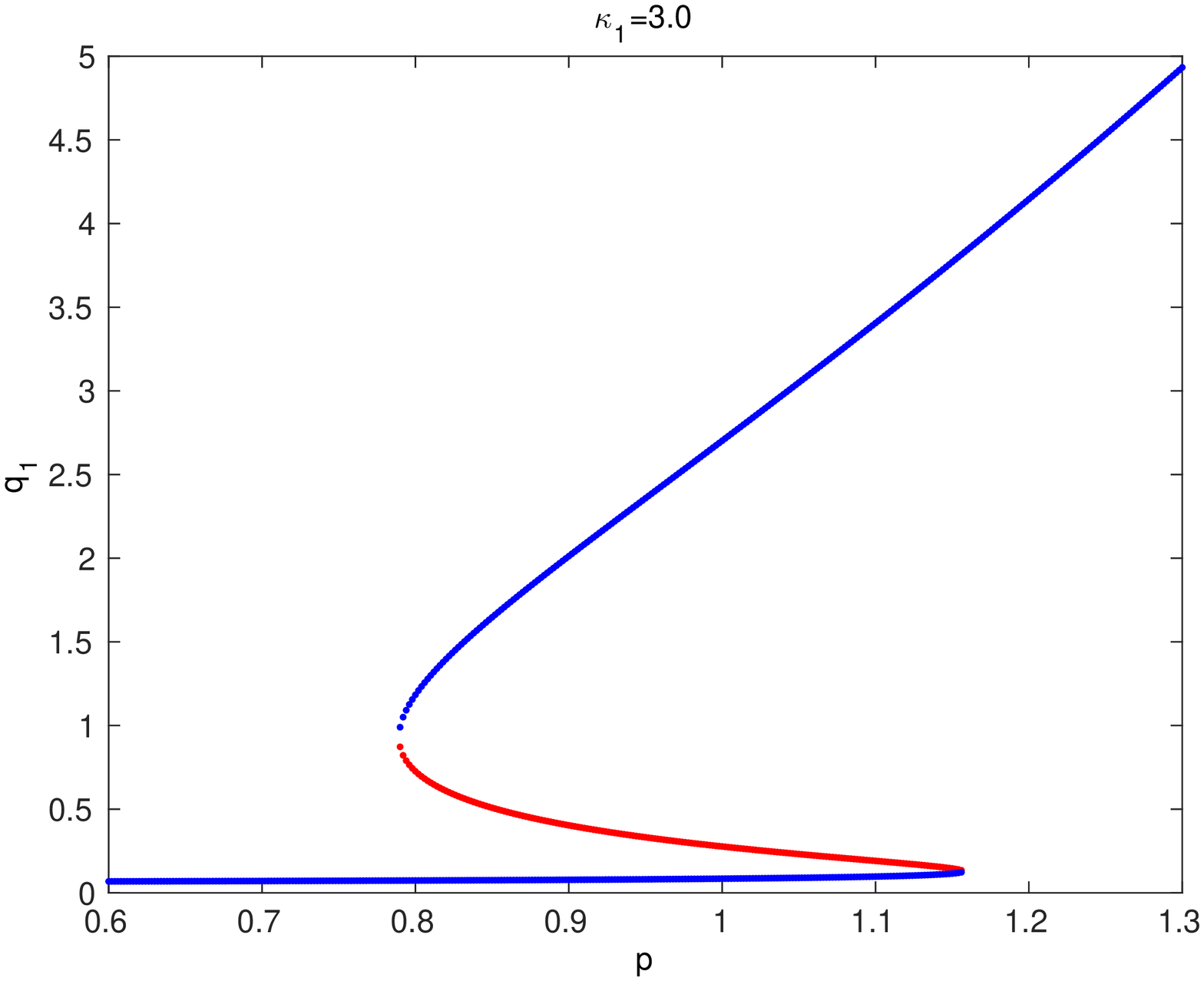} & \includegraphics[scale=0.43]{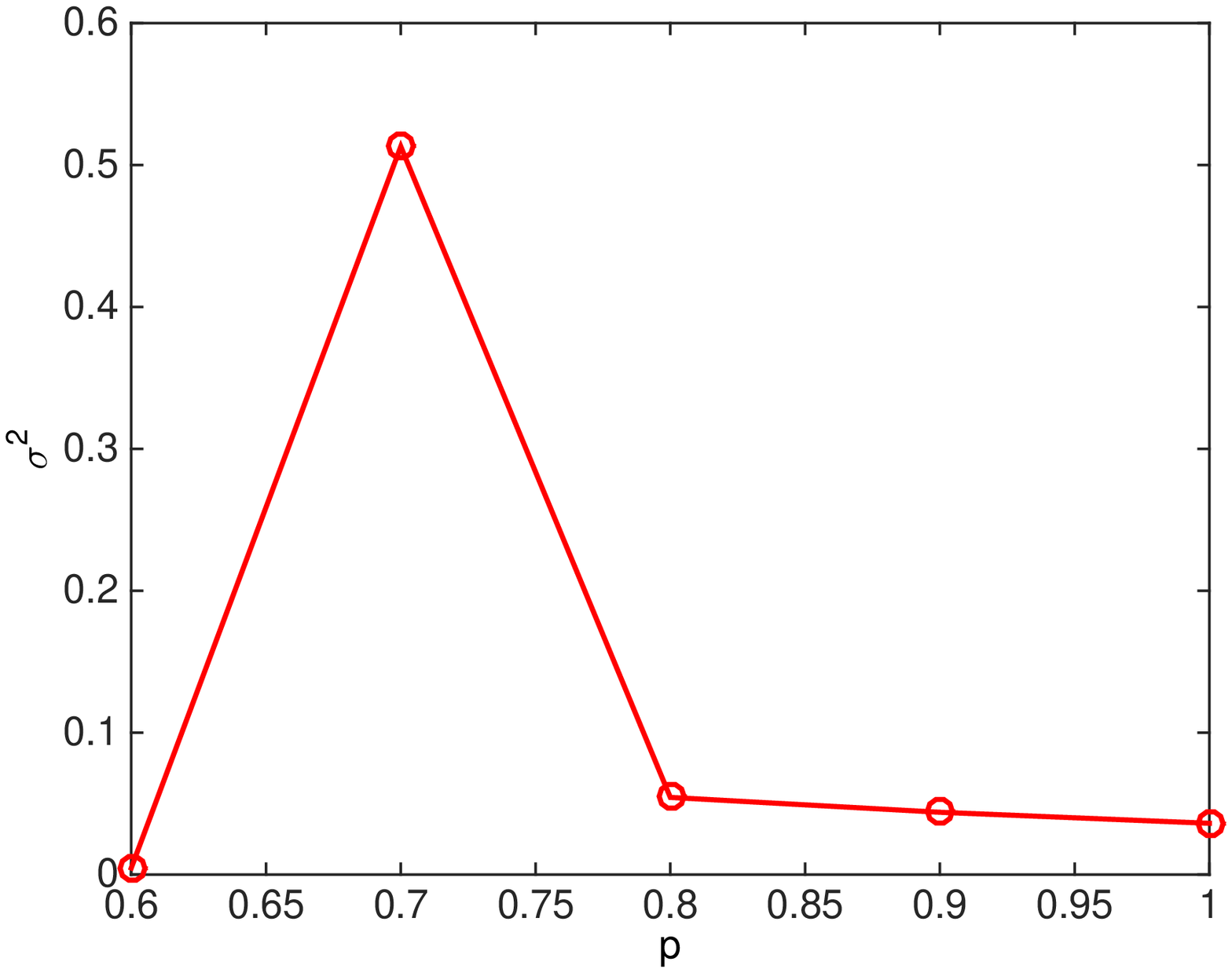}
\end{array}$
\caption{Plot (a): Bifurcation analysis for the SCQSS approximation of the stochastic auto-activation gene regulatory circuit Eqs.
(\ref{eq:qss-sa-q1})-(\ref{eq:qss-sa-q2}), with $\kappa_{1}=3.0$ Parameter values as
given in Table \ref{tab:par-val-selfact}. Plot (b): Simulation results for the variance $\sigma^2=\langle(x_1-\langle x_1\rangle)^2\rangle$, with 
$x_1=X_1(\tau=T)/S$, 
associated to the simulation results shown in Fig. \ref{fig:histselfact}. This plot shows how 
$\sigma^2$ changes as the control parameter, $p$. The maximum of $\sigma^2$ as a function of the control parameter helps us to quantitatively 
determine the corresponding critical value \cite{goldenfeld1992}.}\label{fig:bifdiagselfk3}
\end{center}
\end{figure}

Quantitative comparison between our asymptotic analysis and the simulation results follows the same procedure as in Section \ref{sec:enzyme}, i.e. we 
look at how the variance   aforementioned behaviour regarding unbounded increase of fluctuations close to a bifurcation \cite{goldenfeld1992} is used 
to locate the critical the variance $\sigma^2=\langle(x_1-\langle x_1\rangle)^2\rangle$, with 
$x_1=X_1(\tau=T)/S$ changes as the control parameter varies: the maximum of $\sigma^2$ as a function of the control parameter corresponds to the 
critical value. According to Fig. \ref{fig:bifdiagselfk3}(b), the critical value of $p$, $p_B$, is approximately given by $p_B\simeq 0.7$. Our 
asymptotic analysis (see Fig. \ref{fig:bifdiagselfk3}(b)) predicts that $p_B=0.78$.

\section{Conclusions \& discussion}

By means of the semi-classical quasi-steady state approximation, Section \ref{sec:theory}, we have analysed stochastic effects affecting the onset of bi-stability in cell regulatory systems. Our theory shows that there exists a conserved momentum coordinate associated to each conserved chemical species. In the case of the enzyme-catalysed bistable system, Section \ref{sec:enzyme}, there are three such conserved momenta, associated to each of the conserved chemical species, i.e. Cdh1 and its activating and inhibiting enzymes. For the self-activation gene regulatory network, we have one conserved momentum, corresponding to conservation of the number of binding sites of the gene's promoter region. 

According to the SCQSSA analysis of \cite{alarcon2014}, the maximum rate achieved by an enzymatic reaction, $V_{max}$, predicted by the mean-field theory \cite{keener1998} is renormalised by a factor which equals the value of the (constant) momentum coordinate $p_i$ associated to the conserved enzyme: $V_{max}^{(SC)}=p_{e_{j}}p_iV_{max}$ where $V_{max}^{(SC)}$ is the maximum rate predicted by the SCQSSA. Similarly, we have shown that the mean-field maximum activation rate associated to the auto-activation gene regulatory model, $A_{max}$, is renormalised in the presence of noise by a factor equal to the conserved momentum coordinate corresponding to the number of binding sites in the gene promoter, $p_2$, i.e. $A_{max}^{(SC)}=pp_2A_{max}$, with $A_{max}^{(SC)}$ being the SCQSSA maximum activation rate. As a consequence of this parameter renormalisation, we have shown that variation in the value of the conserved momenta can trigger bifurcations leading to the onset of bistable behaviour beyond the predictions of the mean-field limit, i.e. for values of parameters where the mean-field limit predicts the system to be mono-stable, the SCQSSA predicts bi-stability, and vice versa (see Figs. \ref{fig:bifanTN} and \ref{fig:bifdiagselfact}).

Furthermore, we have established that the value of the constant momenta is actually determined by the probability distribution of the associated conserved chemical species, and, ultimately, by the number of molecules of these species (see Eqs. (\ref{eq:laplaceapprox}) and
(\ref{eq:selfact-p2})-(\ref{eq:selfact-p2-poisson})). Therefore, our theory establishes that the numbers of molecules of the conserved species are
order parameters whose variation should trigger (or cancel) bistable behaviour in the associated systems. This prediction is fully confirmed by direct 
numerical simulation using the stochastic simulation algorithm (see Figs. \ref{fig:ratios}, \ref{fig:p1}, and \ref{fig:histselfact}). Quantitative 
comparison between the predictions of our asymptotic analysis and the simulation results (see Fig. \ref{fig:sigma2enzyme} and 
\ref{fig:bifdiagselfk3}) shows that our theoretical approach slightly underestimates the critical value for the bistable enzyme-regulated system. 
The theoretical prediction for the self-activating gene regulatory network appears to slightly overestimate the critical value. 

Our results allow us to propose a means of controlling cell function. For example, regarding the enzyme-catalysed bistable model analysed in
Section \ref{sec:enzyme}, varying the number of molecules of the three conserved chemical species (Cdh1 and the associated activating and inhibiting enzymes) enables us to lock the system into either of the G$_1$ or the S-G$_2$-M stable fixed points or to drive the system into its bistable regime where random fluctuations will trigger switching between these two states. This could be accomplished by ectopically increasing the synthesis of the corresponding molecule or by targeting the enzymes with enzyme-targeted drugs \cite{robertson2005,singh2012}. Similarly, the dynamics of the self-activating gene regulatory system could be driven into or out of its bistable regime by supplying an inhibitor that irreversibly binds to the promoter region, thus decreasing the effective number of binding sites. 

This result allows us to explore strategies, for example, in the field of combination therapies in cancer treatment. Cellular quiescence is a major factor in resistance to unspecific therapies, such as chemo- and radio-therapy, which target proliferating cells. Bi-stability is central to control cell-cycle progression and to regulate the exit from quiescence, with enzyme catalysis (usually accounted for by (mean-field) Michaelis-Menten, quasi-steady state dynamics) being ubiquitously involved \cite{yao2008,lee2010,bedessem2014,yao2014}. Our findings will allow us to formulate combination strategies in which chemo- or radio-therapy are combined with a strategy aimed at driving cancer cells into proliferation or quiescence depending on the phase of the treatment cycle. Evaluation of the viability and efficiency of such combination requires the formulation of multi-scale models \cite{alarcon2005,guerrero2014a} whose analysis is beyond this scope of this paper, and it is therefore postponed for future work. 

 Our approach differs from previous work, such as Dykman et al.\cite{dykman1994} in a significant aspect, namely, whilst their aim is to 
estimate the rate of noise-induced transition between metastable states in systems exhibiting multi-stability, the purpose of our analysis is to 
ascertain whether noise can alter the multi-stability status of the system. Dykman et al.\cite{dykman1994} do not address such issue.

Eqs. (\ref{eq:hjqss-x1})-(\ref{eq:hjqss-p4}) and (\ref{eq:qss-sa-q1})-(\ref{eq:qss-sa-q2}) are derived from a semi-classical 
approximation of the Master Equation (or its equivalent description in terms or the generating function PDE). This approximation yields a set Hamilton 
equations (Eqs. (8)-(9)) whose solutions are the optimal fluctuation paths and, as such, they describe fluctuation-induced phenomena which cannot be 
accounted for by the mean-field approximation. One of the best known examples of this is exit problems from meta-stable states in noisy systems (e.g. 
extinctions), where the semi-classical approximation provides the optimal escape path from which information such as mean-first passage time or 
waiting time for extinction can be obtained (see, for example, references \cite{elgart2004,khasin2009,assaf2010}). Furthermore, Eqs. 
(\ref{eq:hjqss-x1})-(\ref{eq:hjqss-p4}) and (\ref{eq:qss-sa-q1})-(\ref{eq:qss-sa-q2}) are derived from the general Hamilton equations, Eqs. (8)-(9), 
by means of an approximation based on separation of time scales, not on any mean-field assumption.

A closely related subject to that analysed in this paper is that of noise-induced bifurcations \cite{garciaojalvo1999}. Such phenomenon has been studied in biological systems where the mean-field limit does not predict bistability, such as the so-called enzymatic futile cycles \cite{samoilov2005} where noise associated to the number of enzymes induce bistability. In the absence of this source of noise, the system does not exhibit bistable behaviour. We have not dealt with such noise-induced phenomena in the present paper, in the sense that all the systems analysed in this paper are such that their mean-field limit exhibits bistability. We leave the interesting issue of whether our SCQSSA framework can be used to analyse noise-induced bifurcation phenomena for future research. 

\paragraph*{Acknowledgements.} R.C. and T.A. acknowledge the Spanish Ministry for Science and Innovation (MICINN) for funding MTM2011-29342 and Generalitat de Catalunya for funding under grant 2014SGR1307. R.C. acknowledges AGAUR-Generalitat de Catalunya for funding under its doctoral scholarship programme. P.G. thanks the Wellcome Trust for financial support under grant 098325.


\begin{thebibliography}{83}%
\makeatletter
\providecommand \@ifxundefined [1]{%
 \@ifx{#1\undefined}
}%
\providecommand \@ifnum [1]{%
 \ifnum #1\expandafter \@firstoftwo
 \else \expandafter \@secondoftwo
 \fi
}%
\providecommand \@ifx [1]{%
 \ifx #1\expandafter \@firstoftwo
 \else \expandafter \@secondoftwo
 \fi
}%
\providecommand \natexlab [1]{#1}%
\providecommand \enquote  [1]{``#1''}%
\providecommand \bibnamefont  [1]{#1}%
\providecommand \bibfnamefont [1]{#1}%
\providecommand \citenamefont [1]{#1}%
\providecommand \href@noop [0]{\@secondoftwo}%
\providecommand \href [0]{\begingroup \@sanitize@url \@href}%
\providecommand \@href[1]{\@@startlink{#1}\@@href}%
\providecommand \@@href[1]{\endgroup#1\@@endlink}%
\providecommand \@sanitize@url [0]{\catcode `\\12\catcode `\$12\catcode
  `\&12\catcode `\#12\catcode `\^12\catcode `\_12\catcode `\%12\relax}%
\providecommand \@@startlink[1]{}%
\providecommand \@@endlink[0]{}%
\providecommand \url  [0]{\begingroup\@sanitize@url \@url }%
\providecommand \@url [1]{\endgroup\@href {#1}{\urlprefix }}%
\providecommand \urlprefix  [0]{URL }%
\providecommand \Eprint [0]{\href }%
\providecommand \doibase [0]{http://dx.doi.org/}%
\providecommand \selectlanguage [0]{\@gobble}%
\providecommand \bibinfo  [0]{\@secondoftwo}%
\providecommand \bibfield  [0]{\@secondoftwo}%
\providecommand \translation [1]{[#1]}%
\providecommand \BibitemOpen [0]{}%
\providecommand \bibitemStop [0]{}%
\providecommand \bibitemNoStop [0]{.\EOS\space}%
\providecommand \EOS [0]{\spacefactor3000\relax}%
\providecommand \BibitemShut  [1]{\csname bibitem#1\endcsname}%
\let\auto@bib@innerbib\@empty
\bibitem [{\citenamefont {Kepler}\ and\ \citenamefont
  {Elston}(2001)}]{kepler2001}%
  \BibitemOpen
  \bibfield  {author} {\bibinfo {author} {\bibfnamefont {T.~B.}\ \bibnamefont
  {Kepler}}\ and\ \bibinfo {author} {\bibfnamefont {T.~C.}\ \bibnamefont
  {Elston}},\ }\href@noop {} {\bibfield  {journal} {\bibinfo  {journal}
  {Biophys. J.}\ }\textbf {\bibinfo {volume} {81}},\ \bibinfo {pages} {3116}
  (\bibinfo {year} {2001})}\BibitemShut {NoStop}%
\bibitem [{\citenamefont {Kaern}\ \emph {et~al.}(2005)\citenamefont {Kaern},
  \citenamefont {Elston}, \citenamefont {Blake},\ and\ \citenamefont
  {Collins}}]{kaern2005}%
  \BibitemOpen
  \bibfield  {author} {\bibinfo {author} {\bibfnamefont {M.}~\bibnamefont
  {Kaern}}, \bibinfo {author} {\bibfnamefont {T.~C.}\ \bibnamefont {Elston}},
  \bibinfo {author} {\bibfnamefont {W.~J.}\ \bibnamefont {Blake}}, \ and\
  \bibinfo {author} {\bibfnamefont {J.~J.}\ \bibnamefont {Collins}},\
  }\href@noop {} {\bibfield  {journal} {\bibinfo  {journal} {Nature Rev. Gen.}\
  }\textbf {\bibinfo {volume} {6}},\ \bibinfo {pages} {451} (\bibinfo {year}
  {2005})}\BibitemShut {NoStop}%
\bibitem [{\citenamefont {Maheshri}\ and\ \citenamefont
  {O'Shea}(2007)}]{maheshri2007}%
  \BibitemOpen
  \bibfield  {author} {\bibinfo {author} {\bibfnamefont {N.}~\bibnamefont
  {Maheshri}}\ and\ \bibinfo {author} {\bibfnamefont {E.~K.}\ \bibnamefont
  {O'Shea}},\ }\href@noop {} {\bibfield  {journal} {\bibinfo  {journal} {Annu.
  Rev. Biophys. Biolmol. Struct.}\ }\textbf {\bibinfo {volume} {36}},\ \bibinfo
  {pages} {413} (\bibinfo {year} {2007})}\BibitemShut {NoStop}%
\bibitem [{\citenamefont {Losick}\ and\ \citenamefont
  {Desplan}(2008)}]{losick2008}%
  \BibitemOpen
  \bibfield  {author} {\bibinfo {author} {\bibfnamefont {R.}~\bibnamefont
  {Losick}}\ and\ \bibinfo {author} {\bibfnamefont {C.}~\bibnamefont
  {Desplan}},\ }\href@noop {} {\bibfield  {journal} {\bibinfo  {journal}
  {Science}\ }\textbf {\bibinfo {volume} {320}},\ \bibinfo {pages} {65}
  (\bibinfo {year} {2008})}\BibitemShut {NoStop}%
\bibitem [{\citenamefont {Raj}\ and\ \citenamefont {van
  Oudenaarden}(2008)}]{raj2008}%
  \BibitemOpen
  \bibfield  {author} {\bibinfo {author} {\bibfnamefont {A.}~\bibnamefont
  {Raj}}\ and\ \bibinfo {author} {\bibfnamefont {A.}~\bibnamefont {van
  Oudenaarden}},\ }\href@noop {} {\bibfield  {journal} {\bibinfo  {journal}
  {Cell}\ }\textbf {\bibinfo {volume} {135}},\ \bibinfo {pages} {216} (\bibinfo
  {year} {2008})}\BibitemShut {NoStop}%
\bibitem [{\citenamefont {Cai}, \citenamefont {Dalal},\ and\ \citenamefont
  {Elowitz}(2008)}]{cai2008}%
  \BibitemOpen
  \bibfield  {author} {\bibinfo {author} {\bibfnamefont {L.}~\bibnamefont
  {Cai}}, \bibinfo {author} {\bibfnamefont {C.~K.}\ \bibnamefont {Dalal}}, \
  and\ \bibinfo {author} {\bibfnamefont {M.~B.}\ \bibnamefont {Elowitz}},\
  }\href@noop {} {\bibfield  {journal} {\bibinfo  {journal} {Nature}\ }\textbf
  {\bibinfo {volume} {455}},\ \bibinfo {pages} {485} (\bibinfo {year}
  {2008})}\BibitemShut {NoStop}%
\bibitem [{\citenamefont {Eldar}\ and\ \citenamefont
  {Elowitz}(2010)}]{eldar2010}%
  \BibitemOpen
  \bibfield  {author} {\bibinfo {author} {\bibfnamefont {A.}~\bibnamefont
  {Eldar}}\ and\ \bibinfo {author} {\bibfnamefont {M.~B.}\ \bibnamefont
  {Elowitz}},\ }\href@noop {} {\bibfield  {journal} {\bibinfo  {journal}
  {Nature}\ }\textbf {\bibinfo {volume} {467}},\ \bibinfo {pages} {167}
  (\bibinfo {year} {2010})}\BibitemShut {NoStop}%
\bibitem [{\citenamefont {Kussell}\ and\ \citenamefont
  {Leibler}(2005)}]{kussell2005}%
  \BibitemOpen
  \bibfield  {author} {\bibinfo {author} {\bibfnamefont {E.}~\bibnamefont
  {Kussell}}\ and\ \bibinfo {author} {\bibfnamefont {S.}~\bibnamefont
  {Leibler}},\ }\href@noop {} {\bibfield  {journal} {\bibinfo  {journal}
  {Science}\ }\textbf {\bibinfo {volume} {309}},\ \bibinfo {pages} {2075}
  (\bibinfo {year} {2005})}\BibitemShut {NoStop}%
\bibitem [{\citenamefont {Acar}, \citenamefont {Mettetal},\ and\ \citenamefont
  {van Oudenaarden}(2008)}]{acar2008}%
  \BibitemOpen
  \bibfield  {author} {\bibinfo {author} {\bibfnamefont {M.}~\bibnamefont
  {Acar}}, \bibinfo {author} {\bibfnamefont {J.~T.}\ \bibnamefont {Mettetal}},
  \ and\ \bibinfo {author} {\bibfnamefont {A.}~\bibnamefont {van
  Oudenaarden}},\ }\href@noop {} {\bibfield  {journal} {\bibinfo  {journal}
  {Nature Gen.}\ }\textbf {\bibinfo {volume} {40}},\ \bibinfo {pages} {471}
  (\bibinfo {year} {2008})}\BibitemShut {NoStop}%
\bibitem [{\citenamefont {Guerrero}\ \emph {et~al.}(2015)\citenamefont
  {Guerrero}, \citenamefont {Byrne}, \citenamefont {Maini},\ and\ \citenamefont
  {Alarcon}}]{guerrero2015}%
  \BibitemOpen
  \bibfield  {author} {\bibinfo {author} {\bibfnamefont {P.}~\bibnamefont
  {Guerrero}}, \bibinfo {author} {\bibfnamefont {H.~M.}\ \bibnamefont {Byrne}},
  \bibinfo {author} {\bibfnamefont {P.~K.}\ \bibnamefont {Maini}}, \ and\
  \bibinfo {author} {\bibfnamefont {T.}~\bibnamefont {Alarcon}},\ }\href@noop
  {} {\bibfield  {journal} {\bibinfo  {journal} {J. Math. Biol.}\ ,\ \bibinfo
  {pages} {To appear. DOI: 10.1007/s00285}} (\bibinfo {year}
  {2015})}\BibitemShut {NoStop}%
\bibitem [{\citenamefont {MacArthur}, \citenamefont {Please},\ and\
  \citenamefont {Oreffo}(2008)}]{macarthur2008}%
  \BibitemOpen
  \bibfield  {author} {\bibinfo {author} {\bibfnamefont {B.~D.}\ \bibnamefont
  {MacArthur}}, \bibinfo {author} {\bibfnamefont {C.~P.}\ \bibnamefont
  {Please}}, \ and\ \bibinfo {author} {\bibfnamefont {R.~O.~C.}\ \bibnamefont
  {Oreffo}},\ }\href@noop {} {\bibfield  {journal} {\bibinfo  {journal} {PLoS
  One}\ }\textbf {\bibinfo {volume} {3}},\ \bibinfo {pages} {e3086} (\bibinfo
  {year} {2008})}\BibitemShut {NoStop}%
\bibitem [{\citenamefont {Balazsi}, \citenamefont {van Oudenaarden},\ and\
  \citenamefont {Collins}(2011)}]{balazsi2011}%
  \BibitemOpen
  \bibfield  {author} {\bibinfo {author} {\bibfnamefont {G.}~\bibnamefont
  {Balazsi}}, \bibinfo {author} {\bibfnamefont {A.}~\bibnamefont {van
  Oudenaarden}}, \ and\ \bibinfo {author} {\bibfnamefont {J.~J.}\ \bibnamefont
  {Collins}},\ }\href@noop {} {\bibfield  {journal} {\bibinfo  {journal}
  {Cell}\ }\textbf {\bibinfo {volume} {144}},\ \bibinfo {pages} {910} (\bibinfo
  {year} {2011})}\BibitemShut {NoStop}%
\bibitem [{\citenamefont {Cinquin}\ and\ \citenamefont
  {Demongeot}(2005)}]{cinquin2005}%
  \BibitemOpen
  \bibfield  {author} {\bibinfo {author} {\bibfnamefont {O.}~\bibnamefont
  {Cinquin}}\ and\ \bibinfo {author} {\bibfnamefont {J.}~\bibnamefont
  {Demongeot}},\ }\href@noop {} {\bibfield  {journal} {\bibinfo  {journal} {J.
  theor. Biol.}\ }\textbf {\bibinfo {volume} {233}},\ \bibinfo {pages} {391}
  (\bibinfo {year} {2005})}\BibitemShut {NoStop}%
\bibitem [{\citenamefont {Jaeger}\ and\ \citenamefont
  {Monk}(2014)}]{jaeger2014}%
  \BibitemOpen
  \bibfield  {author} {\bibinfo {author} {\bibfnamefont {J.}~\bibnamefont
  {Jaeger}}\ and\ \bibinfo {author} {\bibfnamefont {N.}~\bibnamefont {Monk}},\
  }\href@noop {} {\bibfield  {journal} {\bibinfo  {journal} {J. Physiol.}\
  }\textbf {\bibinfo {volume} {592}},\ \bibinfo {pages} {2267} (\bibinfo {year}
  {2014})}\BibitemShut {NoStop}%
\bibitem [{\citenamefont {Kauffman}(1993)}]{kauffman1993}%
  \BibitemOpen
  \bibfield  {author} {\bibinfo {author} {\bibfnamefont {S.~A.}\ \bibnamefont
  {Kauffman}},\ }\href@noop {} {\emph {\bibinfo {title} {{The origins of
  order}}}}\ (\bibinfo  {publisher} {Oxford University Press, New York,
  U.S.A.},\ \bibinfo {year} {1993})\BibitemShut {NoStop}%
\bibitem [{\citenamefont {Huang}(2012)}]{huang2012}%
  \BibitemOpen
  \bibfield  {author} {\bibinfo {author} {\bibfnamefont {S.}~\bibnamefont
  {Huang}},\ }\href@noop {} {\bibfield  {journal} {\bibinfo  {journal}
  {BioEssays}\ }\textbf {\bibinfo {volume} {34}},\ \bibinfo {pages} {149}
  (\bibinfo {year} {2012})}\BibitemShut {NoStop}%
\bibitem [{\citenamefont {Tyson}, \citenamefont {Chen},\ and\ \citenamefont
  {Novak}(2003)}]{tyson2003}%
  \BibitemOpen
  \bibfield  {author} {\bibinfo {author} {\bibfnamefont {J.~J.}\ \bibnamefont
  {Tyson}}, \bibinfo {author} {\bibfnamefont {K.~C.}\ \bibnamefont {Chen}}, \
  and\ \bibinfo {author} {\bibfnamefont {B.}~\bibnamefont {Novak}},\
  }\href@noop {} {\bibfield  {journal} {\bibinfo  {journal} {Current Opinion in
  Cell Biology}\ }\textbf {\bibinfo {volume} {15}},\ \bibinfo {pages} {221}
  (\bibinfo {year} {2003})}\BibitemShut {NoStop}%
\bibitem [{\citenamefont {Legewie}, \citenamefont {Bluthgen},\ and\
  \citenamefont {Herzel}(2006)}]{legewie2006}%
  \BibitemOpen
  \bibfield  {author} {\bibinfo {author} {\bibfnamefont {S.}~\bibnamefont
  {Legewie}}, \bibinfo {author} {\bibfnamefont {N.}~\bibnamefont {Bluthgen}}, \
  and\ \bibinfo {author} {\bibfnamefont {H.}~\bibnamefont {Herzel}},\
  }\href@noop {} {\bibfield  {journal} {\bibinfo  {journal} {PLoS Comp. Biol.}\
  }\textbf {\bibinfo {volume} {2}},\ \bibinfo {pages} {e120} (\bibinfo {year}
  {2006})}\BibitemShut {NoStop}%
\bibitem [{\citenamefont {Legewie}, \citenamefont {Bluthgen},\ and\
  \citenamefont {Herzel}(2007)}]{legewie2007}%
  \BibitemOpen
  \bibfield  {author} {\bibinfo {author} {\bibfnamefont {S.}~\bibnamefont
  {Legewie}}, \bibinfo {author} {\bibfnamefont {N.}~\bibnamefont {Bluthgen}}, \
  and\ \bibinfo {author} {\bibfnamefont {H.}~\bibnamefont {Herzel}},\
  }\href@noop {} {\bibfield  {journal} {\bibinfo  {journal} {Biophys. J.}\
  }\textbf {\bibinfo {volume} {93}},\ \bibinfo {pages} {2279} (\bibinfo {year}
  {2007})}\BibitemShut {NoStop}%
\bibitem [{\citenamefont {Kalmar}\ \emph {et~al.}(2009)\citenamefont {Kalmar},
  \citenamefont {Lim}, \citenamefont {Hayward}, \citenamefont {Munoz-Descalzo},
  \citenamefont {Nichols}, \citenamefont {Garcia-Ojalvo},\ and\ \citenamefont
  {Martinez-Arias}}]{kalmar2009}%
  \BibitemOpen
  \bibfield  {author} {\bibinfo {author} {\bibfnamefont {T.}~\bibnamefont
  {Kalmar}}, \bibinfo {author} {\bibfnamefont {C.}~\bibnamefont {Lim}},
  \bibinfo {author} {\bibfnamefont {P.}~\bibnamefont {Hayward}}, \bibinfo
  {author} {\bibfnamefont {S.}~\bibnamefont {Munoz-Descalzo}}, \bibinfo
  {author} {\bibfnamefont {J.}~\bibnamefont {Nichols}}, \bibinfo {author}
  {\bibfnamefont {J.}~\bibnamefont {Garcia-Ojalvo}}, \ and\ \bibinfo {author}
  {\bibfnamefont {A.}~\bibnamefont {Martinez-Arias}},\ }\href@noop {}
  {\bibfield  {journal} {\bibinfo  {journal} {PLoS Biol.}\ }\textbf {\bibinfo
  {volume} {7}},\ \bibinfo {pages} {e1000149} (\bibinfo {year}
  {2009})}\BibitemShut {NoStop}%
\bibitem [{\citenamefont {Ferrel}\ and\ \citenamefont
  {Xiong}(2001)}]{ferrell2001}%
  \BibitemOpen
  \bibfield  {author} {\bibinfo {author} {\bibfnamefont {J.~E.}\ \bibnamefont
  {Ferrel}}\ and\ \bibinfo {author} {\bibfnamefont {W.}~\bibnamefont {Xiong}},\
  }\href@noop {} {\bibfield  {journal} {\bibinfo  {journal} {Chaos}\ }\textbf
  {\bibinfo {volume} {11}},\ \bibinfo {pages} {227} (\bibinfo {year}
  {2001})}\BibitemShut {NoStop}%
\bibitem [{\citenamefont {Tyson}\ and\ \citenamefont
  {Novak}(2001)}]{tyson2001}%
  \BibitemOpen
  \bibfield  {author} {\bibinfo {author} {\bibfnamefont {J.~J.}\ \bibnamefont
  {Tyson}}\ and\ \bibinfo {author} {\bibfnamefont {B.}~\bibnamefont {Novak}},\
  }\href@noop {} {\bibfield  {journal} {\bibinfo  {journal} {J. theor. Biol.}\
  }\textbf {\bibinfo {volume} {210}},\ \bibinfo {pages} {249} (\bibinfo {year}
  {2001})}\BibitemShut {NoStop}%
\bibitem [{\citenamefont {Gerard}\ and\ \citenamefont
  {Goldbeter}(2009)}]{gerard2009}%
  \BibitemOpen
  \bibfield  {author} {\bibinfo {author} {\bibfnamefont {C.}~\bibnamefont
  {Gerard}}\ and\ \bibinfo {author} {\bibfnamefont {A.}~\bibnamefont
  {Goldbeter}},\ }\href@noop {} {\bibfield  {journal} {\bibinfo  {journal}
  {Proc. Natl. Acad. Sci.}\ }\textbf {\bibinfo {volume} {106}},\ \bibinfo
  {pages} {21643} (\bibinfo {year} {2009})}\BibitemShut {NoStop}%
\bibitem [{\citenamefont {Gerard}\ and\ \citenamefont
  {Goldbeter}(2012)}]{gerard2012}%
  \BibitemOpen
  \bibfield  {author} {\bibinfo {author} {\bibfnamefont {C.}~\bibnamefont
  {Gerard}}\ and\ \bibinfo {author} {\bibfnamefont {A.}~\bibnamefont
  {Goldbeter}},\ }\href@noop {} {\bibfield  {journal} {\bibinfo  {journal}
  {Frontiers in Physiology}\ }\textbf {\bibinfo {volume} {3}},\ \bibinfo
  {pages} {413} (\bibinfo {year} {2012})}\BibitemShut {NoStop}%
\bibitem [{\citenamefont {Yao}\ \emph {et~al.}(2012)\citenamefont {Yao},
  \citenamefont {Lee}, \citenamefont {Mori}, \citenamefont {Nevins},\ and\
  \citenamefont {You}}]{yao2008}%
  \BibitemOpen
  \bibfield  {author} {\bibinfo {author} {\bibfnamefont {G.}~\bibnamefont
  {Yao}}, \bibinfo {author} {\bibfnamefont {T.~J.}\ \bibnamefont {Lee}},
  \bibinfo {author} {\bibfnamefont {S.}~\bibnamefont {Mori}}, \bibinfo {author}
  {\bibfnamefont {J.~R.}\ \bibnamefont {Nevins}}, \ and\ \bibinfo {author}
  {\bibfnamefont {L.}~\bibnamefont {You}},\ }\href@noop {} {\bibfield
  {journal} {\bibinfo  {journal} {Nature Cell Biol.}\ }\textbf {\bibinfo
  {volume} {7}},\ \bibinfo {pages} {476} (\bibinfo {year} {2012})}\BibitemShut
  {NoStop}%
\bibitem [{\citenamefont {Yao}(2014)}]{yao2014}%
  \BibitemOpen
  \bibfield  {author} {\bibinfo {author} {\bibfnamefont {G.}~\bibnamefont
  {Yao}},\ }\href@noop {} {\bibfield  {journal} {\bibinfo  {journal} {Interface
  Focus}\ }\textbf {\bibinfo {volume} {4}},\ \bibinfo {pages} {20130074}
  (\bibinfo {year} {2014})}\BibitemShut {NoStop}%
\bibitem [{\citenamefont {Gerard}\ and\ \citenamefont
  {Goldbeter}(2014)}]{gerard2014}%
  \BibitemOpen
  \bibfield  {author} {\bibinfo {author} {\bibfnamefont {C.}~\bibnamefont
  {Gerard}}\ and\ \bibinfo {author} {\bibfnamefont {A.}~\bibnamefont
  {Goldbeter}},\ }\href@noop {} {\bibfield  {journal} {\bibinfo  {journal}
  {Interface Focus}\ }\textbf {\bibinfo {volume} {4}},\ \bibinfo {pages}
  {20130075} (\bibinfo {year} {2014})}\BibitemShut {NoStop}%
\bibitem [{\citenamefont {Bedessem}\ and\ \citenamefont
  {Stephanou}(2014)}]{bedessem2014}%
  \BibitemOpen
  \bibfield  {author} {\bibinfo {author} {\bibfnamefont {B.}~\bibnamefont
  {Bedessem}}\ and\ \bibinfo {author} {\bibfnamefont {A.}~\bibnamefont
  {Stephanou}},\ }\href@noop {} {\bibfield  {journal} {\bibinfo  {journal}
  {Math. Biosci.}\ }\textbf {\bibinfo {volume} {248}},\ \bibinfo {pages} {31}
  (\bibinfo {year} {2014})}\BibitemShut {NoStop}%
\bibitem [{\citenamefont {Keener}\ and\ \citenamefont
  {Sneyd}(1998)}]{keener1998}%
  \BibitemOpen
  \bibfield  {author} {\bibinfo {author} {\bibfnamefont {J.}~\bibnamefont
  {Keener}}\ and\ \bibinfo {author} {\bibfnamefont {J.}~\bibnamefont {Sneyd}},\
  }\href@noop {} {\emph {\bibinfo {title} {{Mathematical physiology}}}}\
  (\bibinfo  {publisher} {Springer-Verlag, New York, NY, USA},\ \bibinfo {year}
  {1998})\BibitemShut {NoStop}%
\bibitem [{\citenamefont {Frigola}\ \emph {et~al.}(2012)\citenamefont
  {Frigola}, \citenamefont {Casanellas}, \citenamefont {Sancho},\ and\
  \citenamefont {Iba\mbox{\~n}es}}]{frigola2012}%
  \BibitemOpen
  \bibfield  {author} {\bibinfo {author} {\bibfnamefont {D.}~\bibnamefont
  {Frigola}}, \bibinfo {author} {\bibfnamefont {L.}~\bibnamefont {Casanellas}},
  \bibinfo {author} {\bibfnamefont {J.~M.}\ \bibnamefont {Sancho}}, \ and\
  \bibinfo {author} {\bibfnamefont {M.}~\bibnamefont {Iba\mbox{\~n}es}},\
  }\href@noop {} {\bibfield  {journal} {\bibinfo  {journal} {PLoS One}\
  }\textbf {\bibinfo {volume} {7}},\ \bibinfo {pages} {e31407} (\bibinfo {year}
  {2012})}\BibitemShut {NoStop}%
\bibitem [{\citenamefont {Garc{\'i}a-Ojalvo}\ and\ \citenamefont
  {Sancho}(1999)}]{garciaojalvo1999}%
  \BibitemOpen
  \bibfield  {author} {\bibinfo {author} {\bibfnamefont {J.}~\bibnamefont
  {Garc{\'i}a-Ojalvo}}\ and\ \bibinfo {author} {\bibfnamefont {J.~M.}\
  \bibnamefont {Sancho}},\ }\href@noop {} {\emph {\bibinfo {title} {{Noise in
  spatially-extended systems}}}}\ (\bibinfo  {publisher} {Springer-Verlag},\
  \bibinfo {year} {1999})\BibitemShut {NoStop}%
\bibitem [{\citenamefont {Samoilov}, \citenamefont {Plyasunov},\ and\
  \citenamefont {Arkin}(2005)}]{samoilov2005}%
  \BibitemOpen
  \bibfield  {author} {\bibinfo {author} {\bibfnamefont {M.}~\bibnamefont
  {Samoilov}}, \bibinfo {author} {\bibfnamefont {S.}~\bibnamefont {Plyasunov}},
  \ and\ \bibinfo {author} {\bibfnamefont {A.~P.}\ \bibnamefont {Arkin}},\
  }\href@noop {} {\bibfield  {journal} {\bibinfo  {journal} {Proc. Natl. Acad.
  Sci.}\ }\textbf {\bibinfo {volume} {102}},\ \bibinfo {pages} {2310} (\bibinfo
  {year} {2005})}\BibitemShut {NoStop}%
\bibitem [{\citenamefont {Rao}\ and\ \citenamefont {Arkin}(2003)}]{rao2003}%
  \BibitemOpen
  \bibfield  {author} {\bibinfo {author} {\bibfnamefont {C.~V.}\ \bibnamefont
  {Rao}}\ and\ \bibinfo {author} {\bibfnamefont {A.~P.}\ \bibnamefont
  {Arkin}},\ }\href@noop {} {\bibfield  {journal} {\bibinfo  {journal} {J.
  Chem. Phys.}\ }\textbf {\bibinfo {volume} {118}},\ \bibinfo {pages} {4999}
  (\bibinfo {year} {2003})}\BibitemShut {NoStop}%
\bibitem [{\citenamefont {Turner}, \citenamefont {Schnell},\ and\ \citenamefont
  {Burrage}(2004)}]{turner2004}%
  \BibitemOpen
  \bibfield  {author} {\bibinfo {author} {\bibfnamefont {T.~E.}\ \bibnamefont
  {Turner}}, \bibinfo {author} {\bibfnamefont {S.}~\bibnamefont {Schnell}}, \
  and\ \bibinfo {author} {\bibfnamefont {K.}~\bibnamefont {Burrage}},\
  }\href@noop {} {\bibfield  {journal} {\bibinfo  {journal} {Comp. Biol.
  Chem.}\ }\textbf {\bibinfo {volume} {28}},\ \bibinfo {pages} {165} (\bibinfo
  {year} {2004})}\BibitemShut {NoStop}%
\bibitem [{\citenamefont {Thomas}, \citenamefont {Straube},\ and\ \citenamefont
  {Grima}(2010)}]{thomas2010}%
  \BibitemOpen
  \bibfield  {author} {\bibinfo {author} {\bibfnamefont {P.}~\bibnamefont
  {Thomas}}, \bibinfo {author} {\bibfnamefont {A.~V.}\ \bibnamefont {Straube}},
  \ and\ \bibinfo {author} {\bibfnamefont {R.}~\bibnamefont {Grima}},\
  }\href@noop {} {\bibfield  {journal} {\bibinfo  {journal} {J. Chem. Phys.}\
  }\textbf {\bibinfo {volume} {133}},\ \bibinfo {pages} {195101} (\bibinfo
  {year} {2010})}\BibitemShut {NoStop}%
\bibitem [{\citenamefont {D{\'o}ka}\ and\ \citenamefont
  {Lente}(2012)}]{doka2012}%
  \BibitemOpen
  \bibfield  {author} {\bibinfo {author} {\bibfnamefont {{\'E}.}~\bibnamefont
  {D{\'o}ka}}\ and\ \bibinfo {author} {\bibfnamefont {G.}~\bibnamefont
  {Lente}},\ }\href@noop {} {\bibfield  {journal} {\bibinfo  {journal} {J.
  Chem. Phys.}\ }\textbf {\bibinfo {volume} {136}},\ \bibinfo {pages} {054111}
  (\bibinfo {year} {2012})}\BibitemShut {NoStop}%
\bibitem [{\citenamefont {Thomas}, \citenamefont {Grima},\ and\ \citenamefont
  {Straube}(2012)}]{thomas2012}%
  \BibitemOpen
  \bibfield  {author} {\bibinfo {author} {\bibfnamefont {P.}~\bibnamefont
  {Thomas}}, \bibinfo {author} {\bibfnamefont {R.}~\bibnamefont {Grima}}, \
  and\ \bibinfo {author} {\bibfnamefont {A.~V.}\ \bibnamefont {Straube}},\
  }\href@noop {} {\bibfield  {journal} {\bibinfo  {journal} {Phys. Rev. E}\
  }\textbf {\bibinfo {volume} {86}},\ \bibinfo {pages} {041110} (\bibinfo
  {year} {2012})}\BibitemShut {NoStop}%
\bibitem [{\citenamefont {Alarc{\'o}n}(2014)}]{alarcon2014}%
  \BibitemOpen
  \bibfield  {author} {\bibinfo {author} {\bibfnamefont {T.}~\bibnamefont
  {Alarc{\'o}n}},\ }\href@noop {} {\bibfield  {journal} {\bibinfo  {journal}
  {J. Phys. Chem.}\ }\textbf {\bibinfo {volume} {140}},\ \bibinfo {pages}
  {184109} (\bibinfo {year} {2014})}\BibitemShut {NoStop}%
\bibitem [{\citenamefont {Bruna}, \citenamefont {Chapman},\ and\ \citenamefont
  {Smith}(2014)}]{bruna2014}%
  \BibitemOpen
  \bibfield  {author} {\bibinfo {author} {\bibfnamefont {M.}~\bibnamefont
  {Bruna}}, \bibinfo {author} {\bibfnamefont {S.~J.}\ \bibnamefont {Chapman}},
  \ and\ \bibinfo {author} {\bibfnamefont {M.~J.}\ \bibnamefont {Smith}},\
  }\href@noop {} {\bibfield  {journal} {\bibinfo  {journal} {J. Chem. Phys.}\
  }\textbf {\bibinfo {volume} {140}},\ \bibinfo {pages} {174107} (\bibinfo
  {year} {2014})}\BibitemShut {NoStop}%
\bibitem [{\citenamefont {Burrage}, \citenamefont {Tian},\ and\ \citenamefont
  {Burrage}(2004)}]{burrage2004}%
  \BibitemOpen
  \bibfield  {author} {\bibinfo {author} {\bibfnamefont {K.}~\bibnamefont
  {Burrage}}, \bibinfo {author} {\bibfnamefont {T.}~\bibnamefont {Tian}}, \
  and\ \bibinfo {author} {\bibfnamefont {P.}~\bibnamefont {Burrage}},\
  }\href@noop {} {\bibfield  {journal} {\bibinfo  {journal} {Progr. Biophys.
  Mol. Biol.}\ }\textbf {\bibinfo {volume} {85}},\ \bibinfo {pages} {217}
  (\bibinfo {year} {2004})}\BibitemShut {NoStop}%
\bibitem [{\citenamefont {Vlachos}(2005)}]{vlachos2005}%
  \BibitemOpen
  \bibfield  {author} {\bibinfo {author} {\bibfnamefont {D.~G.}\ \bibnamefont
  {Vlachos}},\ }\href@noop {} {\bibfield  {journal} {\bibinfo  {journal} {Adv.
  Chem. Phys.}\ }\textbf {\bibinfo {volume} {30}},\ \bibinfo {pages} {1}
  (\bibinfo {year} {2005})}\BibitemShut {NoStop}%
\bibitem [{\citenamefont {MacNamara}, \citenamefont {Burrage},\ and\
  \citenamefont {Sidje}(2008)}]{macnamara2008}%
  \BibitemOpen
  \bibfield  {author} {\bibinfo {author} {\bibfnamefont {S.}~\bibnamefont
  {MacNamara}}, \bibinfo {author} {\bibfnamefont {K.}~\bibnamefont {Burrage}},
  \ and\ \bibinfo {author} {\bibfnamefont {R.~B.}\ \bibnamefont {Sidje}},\
  }\href@noop {} {\bibfield  {journal} {\bibinfo  {journal} {Multiscale Model.
  Simul.}\ }\textbf {\bibinfo {volume} {6}},\ \bibinfo {pages} {1146} (\bibinfo
  {year} {2008})}\BibitemShut {NoStop}%
\bibitem [{\citenamefont {Cao}, \citenamefont {Gillespie},\ and\ \citenamefont
  {Petzold}(2005{\natexlab{a}})}]{cao2005}%
  \BibitemOpen
  \bibfield  {author} {\bibinfo {author} {\bibfnamefont {Y.}~\bibnamefont
  {Cao}}, \bibinfo {author} {\bibfnamefont {D.~T.}\ \bibnamefont {Gillespie}},
  \ and\ \bibinfo {author} {\bibfnamefont {L.~R.}\ \bibnamefont {Petzold}},\
  }\href@noop {} {\bibfield  {journal} {\bibinfo  {journal} {J. Comp. Phys.}\
  }\textbf {\bibinfo {volume} {206}},\ \bibinfo {pages} {395} (\bibinfo {year}
  {2005}{\natexlab{a}})}\BibitemShut {NoStop}%
\bibitem [{\citenamefont {Cao}, \citenamefont {Gillespie},\ and\ \citenamefont
  {Petzold}(2005{\natexlab{b}})}]{cao2005b}%
  \BibitemOpen
  \bibfield  {author} {\bibinfo {author} {\bibfnamefont {Y.}~\bibnamefont
  {Cao}}, \bibinfo {author} {\bibfnamefont {D.~T.}\ \bibnamefont {Gillespie}},
  \ and\ \bibinfo {author} {\bibfnamefont {L.~R.}\ \bibnamefont {Petzold}},\
  }\href@noop {} {\bibfield  {journal} {\bibinfo  {journal} {J. Chem. Phys.}\
  }\textbf {\bibinfo {volume} {122}},\ \bibinfo {pages} {014116} (\bibinfo
  {year} {2005}{\natexlab{b}})}\BibitemShut {NoStop}%
\bibitem [{\citenamefont {Samant}\ and\ \citenamefont
  {Vlachos}(2005)}]{samant2005}%
  \BibitemOpen
  \bibfield  {author} {\bibinfo {author} {\bibfnamefont {A.}~\bibnamefont
  {Samant}}\ and\ \bibinfo {author} {\bibfnamefont {D.~G.}\ \bibnamefont
  {Vlachos}},\ }\href@noop {} {\bibfield  {journal} {\bibinfo  {journal} {J.
  Chem. Phys.}\ }\textbf {\bibinfo {volume} {123}},\ \bibinfo {pages} {144114}
  (\bibinfo {year} {2005})}\BibitemShut {NoStop}%
\bibitem [{\citenamefont {E}, \citenamefont {Liu},\ and\ \citenamefont
  {Vanden-Eijnden}(2007)}]{e2007b}%
  \BibitemOpen
  \bibfield  {author} {\bibinfo {author} {\bibfnamefont {W.}~\bibnamefont {E}},
  \bibinfo {author} {\bibfnamefont {D.}~\bibnamefont {Liu}}, \ and\ \bibinfo
  {author} {\bibfnamefont {E.}~\bibnamefont {Vanden-Eijnden}},\ }\href@noop {}
  {\bibfield  {journal} {\bibinfo  {journal} {J. Comp. Phys.}\ }\textbf
  {\bibinfo {volume} {221}},\ \bibinfo {pages} {158} (\bibinfo {year}
  {2007})}\BibitemShut {NoStop}%
\bibitem [{\citenamefont {Sanft}, \citenamefont {Gillespie},\ and\
  \citenamefont {Petzold}(2011)}]{sanft2011}%
  \BibitemOpen
  \bibfield  {author} {\bibinfo {author} {\bibfnamefont {K.~R.}\ \bibnamefont
  {Sanft}}, \bibinfo {author} {\bibfnamefont {D.~T.}\ \bibnamefont
  {Gillespie}}, \ and\ \bibinfo {author} {\bibfnamefont {L.~R.}\ \bibnamefont
  {Petzold}},\ }\href@noop {} {\bibfield  {journal} {\bibinfo  {journal} {IET
  Syst. Biol.}\ }\textbf {\bibinfo {volume} {5}},\ \bibinfo {pages} {58}
  (\bibinfo {year} {2011})}\BibitemShut {NoStop}%
\bibitem [{\citenamefont {Rathinam}\ \emph {et~al.}(2006)\citenamefont
  {Rathinam}, \citenamefont {Petzold}, \citenamefont {Cao},\ and\ \citenamefont
  {Gillespie}}]{rathinam2003}%
  \BibitemOpen
  \bibfield  {author} {\bibinfo {author} {\bibfnamefont {M.}~\bibnamefont
  {Rathinam}}, \bibinfo {author} {\bibfnamefont {L.~R.}\ \bibnamefont
  {Petzold}}, \bibinfo {author} {\bibfnamefont {Y.}~\bibnamefont {Cao}}, \ and\
  \bibinfo {author} {\bibfnamefont {D.~T.}\ \bibnamefont {Gillespie}},\
  }\href@noop {} {\bibfield  {journal} {\bibinfo  {journal} {J. Chem. Phys.}\
  }\textbf {\bibinfo {volume} {10}},\ \bibinfo {pages} {12784} (\bibinfo {year}
  {2006})}\BibitemShut {NoStop}%
\bibitem [{\citenamefont {Haseltine}\ and\ \citenamefont
  {Rawlings}(2002)}]{haseltine2002}%
  \BibitemOpen
  \bibfield  {author} {\bibinfo {author} {\bibfnamefont {E.~L.}\ \bibnamefont
  {Haseltine}}\ and\ \bibinfo {author} {\bibfnamefont {J.~B.}\ \bibnamefont
  {Rawlings}},\ }\href@noop {} {\bibfield  {journal} {\bibinfo  {journal} {J.
  Chem. Phys.}\ }\textbf {\bibinfo {volume} {117}},\ \bibinfo {pages} {6959}
  (\bibinfo {year} {2002})}\BibitemShut {NoStop}%
\bibitem [{\citenamefont {Salis}\ and\ \citenamefont
  {Kaznessis}(2005)}]{salis2005}%
  \BibitemOpen
  \bibfield  {author} {\bibinfo {author} {\bibfnamefont {H.}~\bibnamefont
  {Salis}}\ and\ \bibinfo {author} {\bibfnamefont {Y.}~\bibnamefont
  {Kaznessis}},\ }\href@noop {} {\bibfield  {journal} {\bibinfo  {journal} {J.
  Chem. Phys.}\ }\textbf {\bibinfo {volume} {122}},\ \bibinfo {pages} {054103}
  (\bibinfo {year} {2005})}\BibitemShut {NoStop}%
\bibitem [{\citenamefont {Assaf}\ and\ \citenamefont
  {Meerson}(2006)}]{assaf2006}%
  \BibitemOpen
  \bibfield  {author} {\bibinfo {author} {\bibfnamefont {M.}~\bibnamefont
  {Assaf}}\ and\ \bibinfo {author} {\bibfnamefont {B.}~\bibnamefont
  {Meerson}},\ }\href@noop {} {\bibfield  {journal} {\bibinfo  {journal} {Phys.
  Rev. E}\ }\textbf {\bibinfo {volume} {74}},\ \bibinfo {pages} {041115}
  (\bibinfo {year} {2006})}\BibitemShut {NoStop}%
\bibitem [{\citenamefont {Newby}\ and\ \citenamefont
  {Chapman}(2013)}]{newby13}%
  \BibitemOpen
  \bibfield  {author} {\bibinfo {author} {\bibfnamefont {J.}~\bibnamefont
  {Newby}}\ and\ \bibinfo {author} {\bibfnamefont {J.}~\bibnamefont
  {Chapman}},\ }\href@noop {} {\bibfield  {journal} {\bibinfo  {journal} {J.
  Math. Biol.}\ ,\ \bibinfo {pages} {1}} (\bibinfo {year} {2013})}\BibitemShut
  {NoStop}%
\bibitem [{\citenamefont {Kang}\ and\ \citenamefont {Kurtz}(2013)}]{kang2013}%
  \BibitemOpen
  \bibfield  {author} {\bibinfo {author} {\bibfnamefont {H.-W.}\ \bibnamefont
  {Kang}}\ and\ \bibinfo {author} {\bibfnamefont {T.~G.}\ \bibnamefont
  {Kurtz}},\ }\href@noop {} {\bibfield  {journal} {\bibinfo  {journal} {The
  Annals of Applied Probability}\ }\textbf {\bibinfo {volume} {23}},\ \bibinfo
  {pages} {529} (\bibinfo {year} {2013})}\BibitemShut {NoStop}%
\bibitem [{\citenamefont {Gillespie}(1976)}]{gillespie1976}%
  \BibitemOpen
  \bibfield  {author} {\bibinfo {author} {\bibfnamefont {D.~T.}\ \bibnamefont
  {Gillespie}},\ }\href@noop {} {\bibfield  {journal} {\bibinfo  {journal} {J.
  Comp. Phys.}\ }\textbf {\bibinfo {volume} {22}},\ \bibinfo {pages} {403}
  (\bibinfo {year} {1976})}\BibitemShut {NoStop}%
\bibitem [{\citenamefont {Assaf}, \citenamefont {Roberts},\ and\ \citenamefont
  {Luthey-Schulten}(2011)}]{assaf2011}%
  \BibitemOpen
  \bibfield  {author} {\bibinfo {author} {\bibfnamefont {M.}~\bibnamefont
  {Assaf}}, \bibinfo {author} {\bibfnamefont {E.}~\bibnamefont {Roberts}}, \
  and\ \bibinfo {author} {\bibfnamefont {Z.}~\bibnamefont {Luthey-Schulten}},\
  }\href@noop {} {\bibfield  {journal} {\bibinfo  {journal} {Phys. Rev. Lett.}\
  }\textbf {\bibinfo {volume} {106}},\ \bibinfo {pages} {248102} (\bibinfo
  {year} {2011})}\BibitemShut {NoStop}%
\bibitem [{\citenamefont {Bressloff}(2014)}]{bressloff2014a}%
  \BibitemOpen
  \bibfield  {author} {\bibinfo {author} {\bibfnamefont {P.~C.}\ \bibnamefont
  {Bressloff}},\ }\href@noop {} {\emph {\bibinfo {title} {{Stochastic processes
  in cell biology.}}}}\ (\bibinfo  {publisher} {Springer-Verlag, Berlin,
  Germany},\ \bibinfo {year} {2014})\BibitemShut {NoStop}%
\bibitem [{\citenamefont {Kampen}(2007)}]{vankampen2007}%
  \BibitemOpen
  \bibfield  {author} {\bibinfo {author} {\bibfnamefont {N.~G.~V.}\
  \bibnamefont {Kampen}},\ }\href@noop {} {\emph {\bibinfo {title} {{Stochastic
  processes in Physics and Chemistry}}}}\ (\bibinfo  {publisher} {Elsevier, The
  Netherlands},\ \bibinfo {year} {2007})\BibitemShut {NoStop}%
\bibitem [{\citenamefont {Kubo}, \citenamefont {Matsuo},\ and\ \citenamefont
  {Kitahara}(1973)}]{kubo1973}%
  \BibitemOpen
  \bibfield  {author} {\bibinfo {author} {\bibfnamefont {R.}~\bibnamefont
  {Kubo}}, \bibinfo {author} {\bibfnamefont {K.}~\bibnamefont {Matsuo}}, \ and\
  \bibinfo {author} {\bibfnamefont {K.}~\bibnamefont {Kitahara}},\ }\href@noop
  {} {\bibfield  {journal} {\bibinfo  {journal} {J. Stat. Phys.}\ }\textbf
  {\bibinfo {volume} {9}},\ \bibinfo {pages} {51} (\bibinfo {year}
  {1973})}\BibitemShut {NoStop}%
\bibitem [{\citenamefont {Alarc{\'o}n}\ and\ \citenamefont
  {Page}(2007)}]{alarcon2007}%
  \BibitemOpen
  \bibfield  {author} {\bibinfo {author} {\bibfnamefont {T.}~\bibnamefont
  {Alarc{\'o}n}}\ and\ \bibinfo {author} {\bibfnamefont {K.~M.}\ \bibnamefont
  {Page}},\ }\href@noop {} {\bibfield  {journal} {\bibinfo  {journal} {J. R.
  Soc. Interface}\ }\textbf {\bibinfo {volume} {4}},\ \bibinfo {pages} {283}
  (\bibinfo {year} {2007})}\BibitemShut {NoStop}%
\bibitem [{\citenamefont {Touchette}(2009)}]{touchette2009}%
  \BibitemOpen
  \bibfield  {author} {\bibinfo {author} {\bibfnamefont {H.}~\bibnamefont
  {Touchette}},\ }\href@noop {} {\bibfield  {journal} {\bibinfo  {journal}
  {Phys. Rep.}\ }\textbf {\bibinfo {volume} {479}},\ \bibinfo {pages} {1}
  (\bibinfo {year} {2009})}\BibitemShut {NoStop}%
\bibitem [{\citenamefont {Doi}(1976)}]{doi1976}%
  \BibitemOpen
  \bibfield  {author} {\bibinfo {author} {\bibfnamefont {M.}~\bibnamefont
  {Doi}},\ }\href@noop {} {\bibfield  {journal} {\bibinfo  {journal} {J. Phys.
  A:Math. Gen.}\ }\textbf {\bibinfo {volume} {9}},\ \bibinfo {pages} {1479}
  (\bibinfo {year} {1976})}\BibitemShut {NoStop}%
\bibitem [{\citenamefont {Peliti}(1985)}]{peliti1985}%
  \BibitemOpen
  \bibfield  {author} {\bibinfo {author} {\bibfnamefont {L.}~\bibnamefont
  {Peliti}},\ }\href@noop {} {\bibfield  {journal} {\bibinfo  {journal} {J.
  Phys. France}\ }\textbf {\bibinfo {volume} {46}},\ \bibinfo {pages} {1469}
  (\bibinfo {year} {1985})}\BibitemShut {NoStop}%
\bibitem [{\citenamefont {Assaf}, \citenamefont {Meerson},\ and\ \citenamefont
  {Sasorov}(2010)}]{assaf2010}%
  \BibitemOpen
  \bibfield  {author} {\bibinfo {author} {\bibfnamefont {M.}~\bibnamefont
  {Assaf}}, \bibinfo {author} {\bibfnamefont {B.}~\bibnamefont {Meerson}}, \
  and\ \bibinfo {author} {\bibfnamefont {P.~V.}\ \bibnamefont {Sasorov}},\
  }\href@noop {} {\bibfield  {journal} {\bibinfo  {journal} {J. Stat. Mech.}\
  ,\ \bibinfo {pages} {P07018}} (\bibinfo {year} {2010})}\BibitemShut {NoStop}%
\bibitem [{\citenamefont {Gonze}, \citenamefont {Halloy},\ and\ \citenamefont
  {Gaspard}(2002)}]{gonze2002}%
  \BibitemOpen
  \bibfield  {author} {\bibinfo {author} {\bibfnamefont {D.}~\bibnamefont
  {Gonze}}, \bibinfo {author} {\bibfnamefont {J.}~\bibnamefont {Halloy}}, \
  and\ \bibinfo {author} {\bibfnamefont {P.}~\bibnamefont {Gaspard}},\
  }\href@noop {} {\bibfield  {journal} {\bibinfo  {journal} {J. Chem. Phys.}\
  }\textbf {\bibinfo {volume} {116}},\ \bibinfo {pages} {10997} (\bibinfo
  {year} {2002})}\BibitemShut {NoStop}%
\bibitem [{\citenamefont {Feynman}\ and\ \citenamefont
  {Hibbs}(2010)}]{feynman2010}%
  \BibitemOpen
  \bibfield  {author} {\bibinfo {author} {\bibfnamefont {R.~P.}\ \bibnamefont
  {Feynman}}\ and\ \bibinfo {author} {\bibfnamefont {A.~R.}\ \bibnamefont
  {Hibbs}},\ }\href@noop {} {\emph {\bibinfo {title} {Quantum Mechanics and
  Path Integrals}}}\ (\bibinfo  {publisher} {Dover Publications, Mineola, NY,
  USA},\ \bibinfo {year} {2010})\BibitemShut {NoStop}%
\bibitem [{\citenamefont {Dickman}\ and\ \citenamefont
  {Vidigal}(2003)}]{dickman2003}%
  \BibitemOpen
  \bibfield  {author} {\bibinfo {author} {\bibfnamefont {R.}~\bibnamefont
  {Dickman}}\ and\ \bibinfo {author} {\bibfnamefont {R.}~\bibnamefont
  {Vidigal}},\ }\href@noop {} {\bibfield  {journal} {\bibinfo  {journal}
  {Brazilian J. Phys.}\ }\textbf {\bibinfo {volume} {33}},\ \bibinfo {pages}
  {73} (\bibinfo {year} {2003})}\BibitemShut {NoStop}%
\bibitem [{\citenamefont {Elgart}\ and\ \citenamefont
  {Kamenev}(2004)}]{elgart2004}%
  \BibitemOpen
  \bibfield  {author} {\bibinfo {author} {\bibfnamefont {V.}~\bibnamefont
  {Elgart}}\ and\ \bibinfo {author} {\bibfnamefont {A.}~\bibnamefont
  {Kamenev}},\ }\href@noop {} {\bibfield  {journal} {\bibinfo  {journal} {Phys.
  Rev. E}\ }\textbf {\bibinfo {volume} {70}},\ \bibinfo {pages} {041106}
  (\bibinfo {year} {2004})}\BibitemShut {NoStop}%
\bibitem [{\citenamefont {T\mbox{\"a}uber}, \citenamefont {Howard},\ and\
  \citenamefont {Vollmayr-Lee}(2005)}]{tauber2005}%
  \BibitemOpen
  \bibfield  {author} {\bibinfo {author} {\bibfnamefont {U.~C.}\ \bibnamefont
  {T\mbox{\"a}uber}}, \bibinfo {author} {\bibfnamefont {M.}~\bibnamefont
  {Howard}}, \ and\ \bibinfo {author} {\bibfnamefont {B.~P.}\ \bibnamefont
  {Vollmayr-Lee}},\ }\href@noop {} {\bibfield  {journal} {\bibinfo  {journal}
  {J. Phys. A: Math. Gen.}\ }\textbf {\bibinfo {volume} {38}},\ \bibinfo
  {pages} {R79} (\bibinfo {year} {2005})}\BibitemShut {NoStop}%
\bibitem [{\citenamefont {Briggs}\ and\ \citenamefont
  {Haldane}(1925)}]{briggs1925}%
  \BibitemOpen
  \bibfield  {author} {\bibinfo {author} {\bibfnamefont {G.~E.}\ \bibnamefont
  {Briggs}}\ and\ \bibinfo {author} {\bibfnamefont {J.~B.~S.}\ \bibnamefont
  {Haldane}},\ }\href@noop {} {\bibfield  {journal} {\bibinfo  {journal}
  {Biochem. J.}\ }\textbf {\bibinfo {volume} {19}},\ \bibinfo {pages} {338}
  (\bibinfo {year} {1925})}\BibitemShut {NoStop}%
\bibitem [{\citenamefont {Guerrero}\ and\ \citenamefont
  {Alarc{\'o}n}(2015)}]{guerrero2014a}%
  \BibitemOpen
  \bibfield  {author} {\bibinfo {author} {\bibfnamefont {P.}~\bibnamefont
  {Guerrero}}\ and\ \bibinfo {author} {\bibfnamefont {T.}~\bibnamefont
  {Alarc{\'o}n}},\ }\href@noop {} {\bibfield  {journal} {\bibinfo  {journal}
  {Math. Model. Nat. Phen.}\ }\textbf {\bibinfo {volume} {10}},\ \bibinfo
  {pages} {64} (\bibinfo {year} {2015})}\BibitemShut {NoStop}%
\bibitem [{\citenamefont {Murray}(1984)}]{murray1984}%
  \BibitemOpen
  \bibfield  {author} {\bibinfo {author} {\bibfnamefont {J.~D.}\ \bibnamefont
  {Murray}},\ }\href@noop {} {\emph {\bibinfo {title} {Asymptotic analysis}}}\
  (\bibinfo  {publisher} {Springer-Verlag, New York, NY, USA},\ \bibinfo {year}
  {1984})\BibitemShut {NoStop}%
\bibitem [{\citenamefont {Ablowitz}\ and\ \citenamefont
  {Fokas}(2003)}]{ablowitz2003}%
  \BibitemOpen
  \bibfield  {author} {\bibinfo {author} {\bibfnamefont {M.~J.}\ \bibnamefont
  {Ablowitz}}\ and\ \bibinfo {author} {\bibfnamefont {A.~S.}\ \bibnamefont
  {Fokas}},\ }\href@noop {} {\emph {\bibinfo {title} {Complex variables.
  Introduction and applications}}}\ (\bibinfo  {publisher} {Cambridge
  University Press, Cambridge, UK},\ \bibinfo {year} {2003})\BibitemShut
  {NoStop}%
\bibitem [{\citenamefont {Goldenfeld}(1992)}]{goldenfeld1992}%
  \BibitemOpen
  \bibfield  {author} {\bibinfo {author} {\bibfnamefont {N.}~\bibnamefont
  {Goldenfeld}},\ }\href@noop {} {\emph {\bibinfo {title} {{Lectures on phase
  transitions and the renormalisation group}}}}\ (\bibinfo  {publisher}
  {Perseus Books Publishing, Reading, Mass., USA},\ \bibinfo {year}
  {1992})\BibitemShut {NoStop}%
\bibitem [{\citenamefont {Weber}\ and\ \citenamefont
  {Buceta}(2013)}]{weber2013}%
  \BibitemOpen
  \bibfield  {author} {\bibinfo {author} {\bibfnamefont {M.}~\bibnamefont
  {Weber}}\ and\ \bibinfo {author} {\bibfnamefont {J.}~\bibnamefont {Buceta}},\
  }\href@noop {} {\bibfield  {journal} {\bibinfo  {journal} {PLoS One}\
  }\textbf {\bibinfo {volume} {8}},\ \bibinfo {pages} {e73487} (\bibinfo {year}
  {2013})}\BibitemShut {NoStop}%
\bibitem [{\citenamefont {Gardner}, \citenamefont {Cantor},\ and\ \citenamefont
  {Collins}(1999)}]{gardner1999}%
  \BibitemOpen
  \bibfield  {author} {\bibinfo {author} {\bibfnamefont {T.~S.}\ \bibnamefont
  {Gardner}}, \bibinfo {author} {\bibfnamefont {C.~R.}\ \bibnamefont {Cantor}},
  \ and\ \bibinfo {author} {\bibfnamefont {J.~J.}\ \bibnamefont {Collins}},\
  }\href@noop {} {\bibfield  {journal} {\bibinfo  {journal} {Nature}\ }\textbf
  {\bibinfo {volume} {403}},\ \bibinfo {pages} {339} (\bibinfo {year}
  {1999})}\BibitemShut {NoStop}%
\bibitem [{\citenamefont {Ozbudak}\ \emph {et~al.}(2004)\citenamefont
  {Ozbudak}, \citenamefont {Thattai}, \citenamefont {Lim}, \citenamefont
  {Shraiman},\ and\ \citenamefont {van Oudenaarden}}]{ozbudak2004}%
  \BibitemOpen
  \bibfield  {author} {\bibinfo {author} {\bibfnamefont {E.~M.}\ \bibnamefont
  {Ozbudak}}, \bibinfo {author} {\bibfnamefont {M.}~\bibnamefont {Thattai}},
  \bibinfo {author} {\bibfnamefont {H.~N.}\ \bibnamefont {Lim}}, \bibinfo
  {author} {\bibfnamefont {B.~I.}\ \bibnamefont {Shraiman}}, \ and\ \bibinfo
  {author} {\bibfnamefont {A.}~\bibnamefont {van Oudenaarden}},\ }\href@noop {}
  {\bibfield  {journal} {\bibinfo  {journal} {Nature}\ }\textbf {\bibinfo
  {volume} {427}},\ \bibinfo {pages} {737} (\bibinfo {year}
  {2004})}\BibitemShut {NoStop}%
\bibitem [{\citenamefont {Lee}\ \emph {et~al.}(2010)\citenamefont {Lee},
  \citenamefont {Yao}, \citenamefont {Bennett}, \citenamefont {Nevins},\ and\
  \citenamefont {You}}]{lee2010}%
  \BibitemOpen
  \bibfield  {author} {\bibinfo {author} {\bibfnamefont {T.~J.}\ \bibnamefont
  {Lee}}, \bibinfo {author} {\bibfnamefont {G.}~\bibnamefont {Yao}}, \bibinfo
  {author} {\bibfnamefont {D.~C.}\ \bibnamefont {Bennett}}, \bibinfo {author}
  {\bibfnamefont {J.~R.}\ \bibnamefont {Nevins}}, \ and\ \bibinfo {author}
  {\bibfnamefont {L.}~\bibnamefont {You}},\ }\href@noop {} {\bibfield
  {journal} {\bibinfo  {journal} {PLoS Biology}\ }\textbf {\bibinfo {volume}
  {8}},\ \bibinfo {pages} {e1000488} (\bibinfo {year} {2010})}\BibitemShut
  {NoStop}%
\bibitem [{\citenamefont {Acar}, \citenamefont {Becksei},\ and\ \citenamefont
  {van Oudenaarden}(2005)}]{acar2005}%
  \BibitemOpen
  \bibfield  {author} {\bibinfo {author} {\bibfnamefont {M.}~\bibnamefont
  {Acar}}, \bibinfo {author} {\bibfnamefont {A.}~\bibnamefont {Becksei}}, \
  and\ \bibinfo {author} {\bibfnamefont {A.}~\bibnamefont {van Oudenaarden}},\
  }\href@noop {} {\bibfield  {journal} {\bibinfo  {journal} {Nature}\ }\textbf
  {\bibinfo {volume} {435}},\ \bibinfo {pages} {228} (\bibinfo {year}
  {2005})}\BibitemShut {NoStop}%
\bibitem [{\citenamefont {Robertson}(2005)}]{robertson2005}%
  \BibitemOpen
  \bibfield  {author} {\bibinfo {author} {\bibfnamefont {J.~G.}\ \bibnamefont
  {Robertson}},\ }\href@noop {} {\bibfield  {journal} {\bibinfo  {journal}
  {Biochemistry}\ }\textbf {\bibinfo {volume} {44}},\ \bibinfo {pages} {5561}
  (\bibinfo {year} {2005})}\BibitemShut {NoStop}%
\bibitem [{\citenamefont {Singh}\ \emph {et~al.}(2012)\citenamefont {Singh},
  \citenamefont {Petter}, \citenamefont {Baillie},\ and\ \citenamefont
  {Whitty}}]{singh2012}%
  \BibitemOpen
  \bibfield  {author} {\bibinfo {author} {\bibfnamefont {J.}~\bibnamefont
  {Singh}}, \bibinfo {author} {\bibfnamefont {R.~C.}\ \bibnamefont {Petter}},
  \bibinfo {author} {\bibfnamefont {T.~A.}\ \bibnamefont {Baillie}}, \ and\
  \bibinfo {author} {\bibfnamefont {A.}~\bibnamefont {Whitty}},\ }\href@noop {}
  {\bibfield  {journal} {\bibinfo  {journal} {Nature Rev. Drug Discovery}\
  }\textbf {\bibinfo {volume} {10}},\ \bibinfo {pages} {307} (\bibinfo {year}
  {2012})}\BibitemShut {NoStop}%
\bibitem [{\citenamefont {Alarc{\'o}n}, \citenamefont {Byrne},\ and\
  \citenamefont {Maini}(2005)}]{alarcon2005}%
  \BibitemOpen
  \bibfield  {author} {\bibinfo {author} {\bibfnamefont {T.}~\bibnamefont
  {Alarc{\'o}n}}, \bibinfo {author} {\bibfnamefont {H.~M.}\ \bibnamefont
  {Byrne}}, \ and\ \bibinfo {author} {\bibfnamefont {P.~K.}\ \bibnamefont
  {Maini}},\ }\href@noop {} {\bibfield  {journal} {\bibinfo  {journal}
  {Multiscale Model. Sim.}\ }\textbf {\bibinfo {volume} {3}},\ \bibinfo {pages}
  {440} (\bibinfo {year} {2005})}\BibitemShut {NoStop}%
\bibitem [{\citenamefont {Dykman}\ \emph {et~al.}(1994)\citenamefont {Dykman},
  \citenamefont {Mori}, \citenamefont {Ross},\ and\ \citenamefont
  {Hunt}}]{dykman1994}%
  \BibitemOpen
  \bibfield  {author} {\bibinfo {author} {\bibfnamefont {M.~I.}\ \bibnamefont
  {Dykman}}, \bibinfo {author} {\bibfnamefont {E.}~\bibnamefont {Mori}},
  \bibinfo {author} {\bibfnamefont {J.}~\bibnamefont {Ross}}, \ and\ \bibinfo
  {author} {\bibfnamefont {P.~M.}\ \bibnamefont {Hunt}},\ }\href@noop {}
  {\bibfield  {journal} {\bibinfo  {journal} {J. Chem. Phys.}\ }\textbf
  {\bibinfo {volume} {100}},\ \bibinfo {pages} {5735} (\bibinfo {year}
  {1994})}\BibitemShut {NoStop}%
\bibitem [{\citenamefont {Khasin}\ and\ \citenamefont
  {Dykman}(2009)}]{khasin2009}%
  \BibitemOpen
  \bibfield  {author} {\bibinfo {author} {\bibfnamefont {M.}~\bibnamefont
  {Khasin}}\ and\ \bibinfo {author} {\bibfnamefont {M.~I.}\ \bibnamefont
  {Dykman}},\ }\href@noop {} {\bibfield  {journal} {\bibinfo  {journal} {Phys.
  Rev. Lett.}\ }\textbf {\bibinfo {volume} {103}},\ \bibinfo {pages} {068101}
  (\bibinfo {year} {2009})}\BibitemShut {NoStop}%
\end{thebibliography}
\end{document}